\documentclass[aps,twocolumn,nofootinbib,groupedaddress,amsfonts,floatfix]{revtex4-1} 
\usepackage{graphicx,amsmath,amssymb,amstext}
\usepackage{amssymb,amsbsy,amsfonts,amsthm,color}

\usepackage{epsfig}
\usepackage{graphicx}
\usepackage{subfigure}
\usepackage{hyperref}
\usepackage{cancel}
\usepackage{multirow}

\graphicspath{{Figures/}}

\usepackage{color}
\usepackage[dvipsnames]{xcolor}

\newcommand{\Beq}{\begin{eqnarray}}
\newcommand{\Eeq}{\end{eqnarray}}

\newcommand{\nn}{\nonumber \\}

\newcommand{\mpl}{M_{\mbox{\tiny Pl}}}

\newcommand{\eqn}[1]{Eqn. (\ref{#1})}
\renewcommand{\eqref}[1]{Eqn. (\ref{#1})}

\begin{document}
{\hfill KCL-PH-TH/2019-84}
\title{The Effects of Potential Shape on Inhomogeneous Inflation}

\author{Josu C. Aurrekoetxea${^a}$}
\email{j.c.aurrekoetxea@gmail.com}
\author{Katy Clough ${^b}$}
\email{katy.clough@physics.ox.ac.uk}
\author{Raphael Flauger ${^c}$}
\email{flauger@physics.ucsd.edu }
\author{Eugene A. Lim ${^a}$\vspace{0.2cm}}
\email{eugene.a.lim@gmail.com}

\affiliation{${^a}$Theoretical Particle Physics and Cosmology Group, Physics Department,
Kings College London, Strand, London WC2R 2LS, United Kingdom}
\affiliation{${^b}$Astrophysics, University of Oxford, DWB, Keble Road, Oxford OX1 3RH, UK}
\affiliation{${^c}$ UC San Diego, Department of Physics, 9500 Gilman Rd, La Jolla, CA, 92093, USA}
\begin{abstract}
We study the robustness of single-field inflation against inhomogeneities.  We derive a simple analytic criterion on the shape of the potential for successful inflation in the presence of inhomogeneities, and demonstrate its validity using full 3+1 dimensional numerical relativity simulations on several classes of popular models of single-field inflation. 
  We find that models with convex potentials are more robust to inhomogeneities than those with concave potentials, and that concave potentials that vary on super-Planckian scales are significantly more robust than those that vary on sub-Planckian scales.

\end{abstract}
\pacs{}
\maketitle

\section{Introduction} \label{sect:intro}

Cosmic inflation \cite{Guth:1980zm,Linde:1981mu,Albrecht:1982wi,Starobinsky:1980te} is the leading paradigm for the early universe. It posits an early period of accelerated expansion in order to dynamically explain the current homogeneous and  spatially flat state of the universe. One of the most remarkable successes of the paradigm is the observational confirmation of some of its key predictions, a nearly scale-invariant and Gaussian spectrum of primordial perturbations \cite{Akrami:2018odb}.
$•$

Inflation was introduced as a solution to several problems of standard big-bang cosmology~\cite{Guth:1980zm}. One of these problems is the horizon problem. However, inflation can only constitute a solution to the horizon problem, if it does not have a horizon problem of its own. So it is natural to ask what came before inflation and how it began. There is certainly no guarantee that the universe was semi-classical at the time inflation began, and it is difficult to talk about the beginning of inflation in complete generality. To make progress, we make the simplifying assumption that at the time inflation began the universe was already described by general relativity minimally coupled to a scalar field, the inflaton.

In this case, the space of initial conditions for inflation is parameterized by the degrees of freedom of the spatial metric and its conjugate momentum, and the corresponding degrees of freedom of the inflaton. For each degree of freedom, we are free to specify its configuration on the initial Cauchy hypersurface. It is then natural to ask for which initial data inflation will be successful. By ``successful'' we mean that some region of the initial hypersurface undergoes accelerated expansion for $60$ $e$-foldings or more. Whether inflation is successful depends both on the dynamics of the inflaton model and the initial conditions, as well as the interplay between them.  

In this work, we will explore one particular aspect of this interplay, the effect of the initial amplitude of the inhomogeneities in the inflaton field (and their associated metric perturbations) on different models of inflation. 

Inflationary models are broadly classified as ``concave'' and ``convex'' models, depending on the shape of their potentials. 
We propose an analytic criterion as a diagnostic for whether inflation will be successful for a given potential. We test this criterion using full 3+1 numerical general relativity solutions and show that convex models are more robust to inhomogeneities than concave models. Furthermore, we show that for concave potentials the scale in field space over which the potential varies appreciably plays an important role. Finally, we will argue that for some potentials there exists a bound on the initial mean value of the inflaton field, beyond which inflation will be successful regardless of the amplitude of the inhomogeneities.

\section{Initial conditions and models} \label{sect:param}

The problem of initial conditions for inflation has been studied extensively using analytic and semi-analytic methods~\cite{Gibbons:1977mu,Hawking:1981fz,Wald:1983ky,Starobinsky:1982mr,Barrow:1984zz,Albrecht:1984qt,Barrow:1985,Gibbons:1986xk,Jensen:1986nf,Hawking:1987bi,Penrose:1988mg,Muller:1989rp,Kitada:1991ih,Kitada:1992uh,Bruni:1994cv,Maleknejad:2012as,Gibbons:2006pa,Boucher:2011zj,Bruni:2001pc,Muller:1987hp,Barrow:1989wp,Bicak:1997ne,Capozziello:1998dq,Vachaspati:1998dy,Barrow:1987ia,Barrow:1986yf,Polyakov:2009nq,Marolf:2010nz,Tsamis:1992sx,Brandenberger:2002sk,Geshnizjani:2003cn,Marozzi:2012tp,Brandenberger:1990wu,Carroll:2010aj,Corichi:2010zp,Schiffrin:2012zf,Remmen:2013eja,Corichi:2013kua,Mukhanov:2014uwa,Remmen:2014mia,Berezhiani:2015ola,Kleban:2016sqm,
Marsh:2018fsu,Finn:2018krt,Bloomfield:2019rbs}, as well as numerically
~\cite{Albrecht:1985yf,Albrecht:1986pi,KurkiSuonio:1987pq,Feldman:1989hh,Brandenberger:1988ir,Goldwirth:1989pr,Goldwirth:1989vz,Brandenberger:1990xu,Laguna:1991zs,Goldwirth:1991rj,KurkiSuonio:1993fg,Easther:2014zga,East:2015ggf,Braden:2016tjn,Alho:2011zz,  Alho:2013vva} (see~\cite{Brandenberger:2016uzh} for a short review).

Recently it has become possible to use numerical relativity codes to evolve different initial configurations in the time domain even in the regime in which black holes form \cite{East:2015ggf,Clough:2016ymm}, allowing for a fully non-perturbative investigation of the field dynamics in response to the initial conditions. 
This work was limited to a small number of ``typical'' models, and quantified their success for different choices of parameterized initial inhomogeneities. 

One natural way to extend these investigations is to expand the classes of inhomogeneities, which was initiated in~\cite{Clough:2017efm}.
Another interesting direction is to expand the classes of inflationary models under investigation, which we will do in the present work. 

In this section we summarize the key features of the space of initial conditions on which we focus, as well as the larger class of inflationary  models.

\subsection{The space of initial conditions}

We decompose the spacetime metric using the standard ADM decomposition \cite{PhysRev.116.1322},
\begin{equation}
ds^2=-\alpha^2\,dt^2+\gamma_{ij}(dx^i + \beta^i\,dt)(dx^j + \beta^j\,dt)\,.
\end{equation}
Here $\gamma_{ij}$ is the 3-metric on the spatial hypersurface, while $\alpha$ and $\beta^i$ are the lapse and shift. We are free to choose the initial Cauchy hypersurface. The metric initial conditions are then fully specified by a choice of $\gamma_{ij}$ and the extrinsic curvature $K_{ij}$ 
at each point in the spatial domain.
The extrinsic curvature can further be decomposed into the expansion $K = \gamma^{ij}K_{ij}$ and trace-free tensor components $A_{ij}$,
\begin{equation}
K_{ij} = \frac{1}{3}K\gamma_{ij}+A_{ij}~.
\end{equation}
In the perturbative limit, the transverse part of $A_{ij}$ represents ``gravitational wave'' modes, though we emphasize that in the non-perturbative limit they are not solutions to a linear wave equation.  In our sign convention, $K<0$ denotes a locally expanding spacetime. 

Meanwhile, the space of initial conditions for the inflaton $\phi$ is given by the value of the field and its canonical momentum 
\begin{equation}
	\Pi\equiv\alpha^{-1}(\dot{\phi} - \beta^i\partial_i \phi)\,,
\end{equation} 
at each point.

A more subtle aspect of the initial conditions is the choice of spatial boundary conditions. In what follows we will assume periodic boundary conditions for the spatial domain, which imposes a $T^3$ topology on the space. Alternatively, this can be thought of as imposing a scale of homogeneity on the initial conditions with the Universe made up of many (inhomogeneous) boxes of size $L$. One can always make $L$ larger, thus increasing the scale of homogeneity relative to our patch of the Universe, and it is usually considered that taking $L$ to be greater than the initial Hubble scale of inflation is a sufficiently conservative approach. Other topologies, in particular those that can support a positive-definite $^{(3)}R$ can lead to different conclusions~\cite{Barrow:1985,Kleban:2016sqm}, so that this is a choice that should be made explicit.

In summary, the space of initial conditions for single-field inflation consists of the values of the variables $(\gamma_{ij},K_{ij},\phi,\Pi)$ on the initial hypersurface. Their values are not completely independent because they are subject to the Hamiltonian and momentum constraints (see \eqref{eq:HamiltonianConstraint}).

The constraints are a set of four non-linear coupled partial differential equations, which are non-trivial to solve for a general matter distribution (see e.g. \cite{Alcubierre:1138167,Baumgarte:1998te}). To simplify this task, some variables are often set to zero, severely restricting the available space of initial conditions. In Refs. \cite{East:2015ggf, Clough:2016ymm, Clough:2017efm}, $\gamma_{ij}$ is, for example, assumed to be conformally flat $\gamma_{ij} \equiv \chi^{-1}\delta_{ij}$ where $\chi$ is a conformal factor.   In addition, Refs. \cite{East:2015ggf, Clough:2016ymm} set the trace-free part of the extrinsic curvature to zero $A_{ij}=0$.  Combined with the additional simplifying conditions that $\Pi=0$ and the expansion rate is spatially constant $K=\mathrm{const}<0$ (i.e. uniformly expanding), the momentum constraint is then trivially satisfied. The parameter space in this case is then just the scalar configuration $\phi({\bf x})$, with the value of $K$ imposed by an integrability condition in the case of periodic boundary conditions (see \cite{Bentivegna:2013xna}). The Hamiltonian constraint then determines the conformal factor, $\chi$. 

Two more general classes of deviations from homogeneity have been explored.  First, in \cite{Clough:2016ymm}, we explored a special case where $K({\bf x}) = -C\phi({\bf x}) + K_0$ where $C>0$  is a free parameter, and the value of $K_0$ is set by integrability on the periodic domain. This \emph{Ansatz} keeps the momentum constraint trivial but allows us to explore initial conditions which mix regions of local expansion ($K<0$) and contraction ($K>0$). We found that, as long as the initial hypersurface is expanding on average,  $\langle K \rangle <0$, inflation will occur in some patch, even if part of the spacetime collapses. Second, in \cite{Clough:2017efm}, we studied non-zero transverse modes $A^{TT}_{ij}\neq 0$. We found that the amplitude of the scalar perturbations remained the main driver of inflationary success, with the tensor modes generally reducing the number of $e$-folds, but not causing failure in isolation even at high relative energy densities.

It is clear from this discussion that a large space of initial conditions remains to be explored. As we mentioned before, we will not pursue this direction here and reserve it for future work. We instead consider initial conditions such that $\tilde \gamma_{ij}=\delta_{ij}$, $\Pi=0$, $A_{ij}=0$ and  $K=\mathrm{const}<0$, and we expand the classes of inflationary models. 

\subsection{The space of models}\label{sec:models}
Inflation a priori only predicts that the observed primordial spectrum of density perturbations should be nearly scale-invariant but does not predict the sign of the departure from scale invariance. The space of inflationary models is vast, encompassing a diverse variety of single field and multi-field models with many mechanisms (see e.g. \cite{Martin:2013tda}). Without the guidance of some fundamental theory of inflation, models which are not already ruled out by observations are in principle all valid. Nevertheless, useful classifications of models can be made, such as categorizing the models in terms of their energy scale, or the field range. As one might expect, we showed in \cite{Clough:2016ymm} that small field inflation is generally less robust than large field inflation.

In this work, we will rely on a slightly more refined classification to guide our choice of inflationary models. In the slow-roll approximation, we know that the slow-roll parameter $\epsilon=-\dot{H}/H^2$ obeys the differential equation
\begin{equation}\label{eq:diffepsilon}
\frac{d \ln \epsilon}{d \mathcal{N}}=\left(n_{\mathrm{s}}(\mathcal{N})-1\right)+2 \epsilon\,.
\end{equation}
Suppose the observed sign and magnitude of the departure from scale-invariance are not mere accidents but arise because the scalar spectral index has the functional form~\cite{Mukhanov:2013tua,Roest:2013fha}\footnote{There is no guarantee that this is the case, but models that deviate from this will require additional small parameters to account for the near scale-invariance.}
\begin{equation}
n_{\mathrm{s}}(\mathcal{N})-1=-\frac{p+1}{\mathcal{N}}\,,
\end{equation}
where $\mathcal{N}$ is the number of $e$-folds (counted from the end of inflation), and $p$ is number of order unity. The most general solution to the differential equation~(\ref{eq:diffepsilon}) is then given by 
\begin{equation}
\epsilon(\mathcal{N})=\frac{p}{2 \mathcal{N}} \frac{1}{1 \pm\left(\mathcal{N} / \mathcal{N}_{\mathrm{eq}}\right)^{p}}\,,
\end{equation}
where $\mathcal{N}_{\mathrm{eq}}$ is an integration constant. Unless there are additional hierarchies, we expect it to be of order unity. 

If we further assume that we observe modes at a typical moment so that either the first or the second term in the denominator dominate, we are left with two solutions that are compatible with current data
\begin{equation}
\epsilon(\mathcal{N})=\frac{p}{2 \mathcal{N}}\qquad\text{and}\qquad \epsilon(\mathcal{N})=\frac{p}{2 \mathcal{N}}\left(\frac{\mathcal{N}_{\mathrm{eq}}}{\mathcal{N}}\right)^{p}\,.
\end{equation}
Because $p<0$ is disfavored by data, we will restrict our attention to $p>0$. 

So far this is general and makes no mention of a potential. If we assume that the dynamics is governed by a single scalar field with canonical kinetic term
\begin{equation}
L_\phi=-\frac{1}{2}g^{\mu\nu}\partial_\mu\phi\partial_\nu\phi - V(\phi)\,,
\end{equation}
we can reconstruct the potential from the equations\footnote{We will use the Planck mass $\mpl^2=\hbar c/G$ throughout the paper.  }
\begin{equation}
\frac{d \phi}{d \mathcal{N}}=\frac{\mpl^2}{8\pi} \frac{V^{\prime}}{V}\qquad\text{and}\qquad\left(\frac{d \phi}{d \mathcal{N}}\right)^{2}=\frac{ \epsilon \mpl^{2}}{4\pi}\,.
\end{equation}
One finds that the first class corresponds to monomial (or power law) potentials
\begin{equation}
V(\phi)=\lambda \mpl^{4-2p}\phi^{2p}\,. \label{eqn:monomial_pot}
\end{equation}
For the second class of models the potential during inflation is well approximated by
\begin{equation}\label{eq:ht}
V(\phi) \simeq \Lambda^4\left[1-\left(\frac{\phi} { \mu_n}\right)^{n}\right]\,,
\end{equation}
where $n=2 p /(p-1)$ provided $p>1$, and by 
\begin{equation}\label{eq:bi}
V(\phi) \simeq \Lambda^4\left[1-\left(\frac{\mu_n}{ \phi}\right)^{n}\right]\,,
\end{equation}
with $n=2 p /(1-p)$ provided $0<p<1$. For potentials of the form~(\ref{eq:ht}) inflation occurs as the inflaton rolls off a hilltop, and we will sometimes refer to them as hilltop models. Similarly, for the form~(\ref{eq:bi}) inflation occurs as the inflaton rolls off a plateau, and we will sometimes refer to them as plateau models. For $p\neq 1$ the departure from the plateau is described by a power law. The case $p=1$ is special and the departure becomes exponential
\begin{equation}\label{eq:explateau}
V(\phi) \simeq \Lambda^4\left[1-e^{-\phi/\mu}\right]\,.
\end{equation}

We see that the hilltop and plateau models involve an additional scale, denoted $\mu$ in the exponential models and $\mu_n$ in the power law models, which describes the distance in field space over which the plateau is approached. As we will see, this scale plays an important role for the robustness against inhomogeneities. Roughly, we will see that models are robust if this scale is Planckian or super-Planckian, and are susceptible to inhomogeneities if this scale is sub-Planckian. Incidentally, this is closely tied to the question whether gravitational waves from these models are detectable with upcoming CMB experiments, which will be capable of detecting gravitational wave signals from models with a super-Planckian characteristic scale~\cite{Abazajian:2016yjj}.

\section{An analytic criterion for robustness} \label{sect:criterion}

In this section, we will derive a simple analytic criterion that allows us to infer the robustness of a given single-field model against inhomogeneities. From~\cite{East:2015ggf,Clough:2016ymm,Clough:2017efm} we know that the amplitude of the inhomogeneity plays an important role. At fixed energy density in the gradients, the amplitude is generically larger if only one or a few modes are excited than if the energy density is distributed over a large number of modes. So in what follows we will assume that the inhomogeneous initial conditions for the inflaton field are given by a superposition of modes in the 3 spatial directions \cite{East:2015ggf,Clough:2016ymm,Clough:2017efm}
\begin{equation}\label{eq:fieldinitial}
\phi(t=0,\mathbf{x}) = \phi_0 + \frac{\Delta\phi}
{3}\left(\cos kx + \cos ky + \cos kz\right)\,,
\end{equation}
with vanishing canonical momentum
\begin{equation}
\Pi(t=0,\mathbf{x}) = 0~, \label{eqn:kinetic_initial}
\end{equation}
where $k=2\pi N/L$ is the wavenumber associated with the inhomogeneity, $N=1,2,\dots$ is an integer, and we set $L$ to be the Hubble length  $H_i^{-1}$ in the absence of inhomogeneities
\begin{equation}\label{eq:length}
L=\frac{3M_\mathrm{Pl}}{\sqrt{24\pi V(\phi_0)}}\,. 
\end{equation}
If we assume that the spatial metric is initially conformally flat, $\gamma_{ij}=\chi^{-1}\delta_{ij}$, and $K=\mathrm{const}<0$ as discussed in Sec. \ref{sect:param}, then the scalar field dynamics \emph{near the initial hyperslice} is approximately described by the Klein-Gordon equation
\begin{equation}\label{eq:kleingordon}
\ddot{\phi}\approx \nabla^2\phi-\frac{dV(\phi)}{d\phi}\,,
\end{equation}
where the subdominant friction term has been ignored since initially $\dot\phi \sim 0$. Without loss of generality, we assume that the inflaton rolls down the potential to the reheating minimum $\phi_{\mathrm{reh}}$ in the positive $\phi$ direction. Consider the dynamics of the inflaton at the point of maximum amplitude $\phi_{\rm max}$ -- this is the field value closest to the reheating minimum and thus the point at which the inflaton is most at risk of falling into the reheating minimum and ending inflation. Our question is then whether the inflaton field at this point will initially be ``pulled back'' towards $\phi_0$ to safety, or whether the gradient of the potential $dV/d\phi<0$ will drag the inflaton into the minimum, ending inflation early or preventing it from taking place altogether.  For compactness, we will sometimes refer to the former as ``pull-back'' and the latter as ``drag-down''. 

\begin{figure}[tb]
\begin{center}
{\includegraphics[width=\columnwidth]{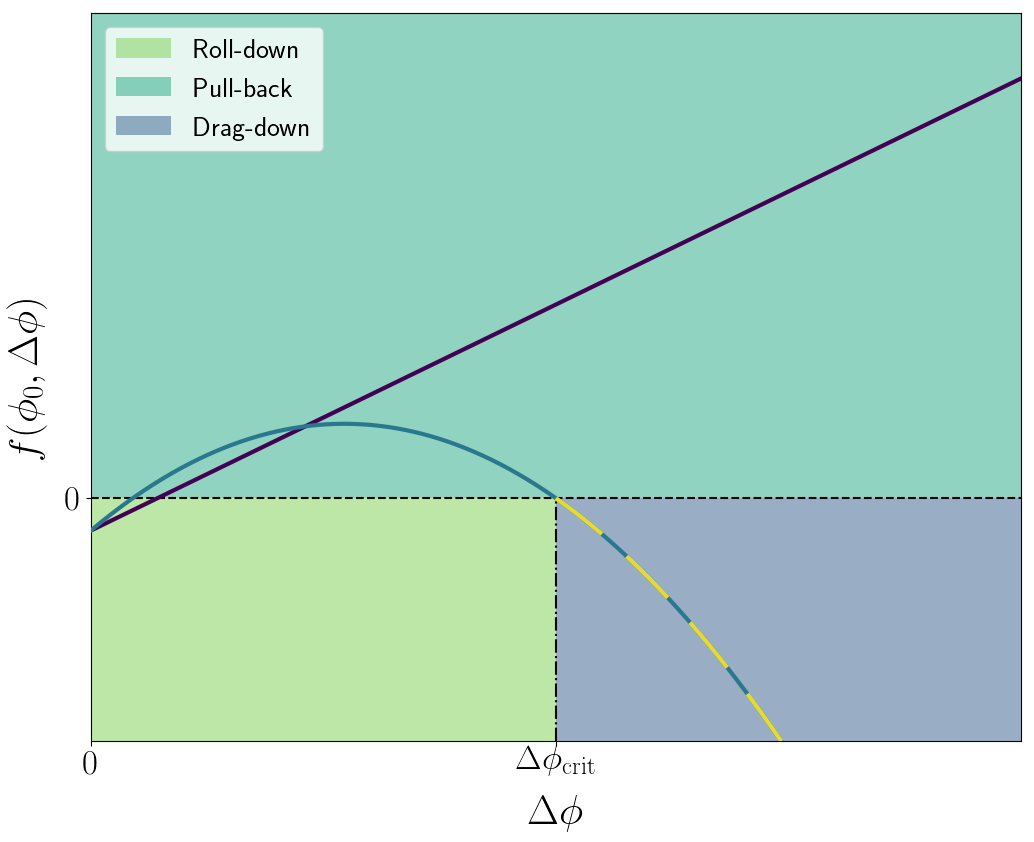}}
\caption{Sketch of $f(\phi_0,\Delta\phi)$ for a concave and convex model: The main difference between models (different solid lines) is whether $f$ has a maximum and there exists a ``drag-down'' region where $f<0$. Away from the trivial $\Delta \phi=0$ homogenous point,  convex models do not have such maximum and always stay within the $f>0$ region where the field is pulled back. Concave models however, can have a turning point followed by a zero crossing and values of $\Delta\phi$ for which $f<0$ (dashed line) and hence the field is dragged towards the minimum, ending inflation.
}
 \label{fig:fFunction}
\end{center}
\end{figure}

Initially, $\phi_{\mathrm{max}}=\phi_0+\Delta\phi$. Using \eqn{eq:kleingordon} and \eqn{eq:fieldinitial}, its initial evolution is given by
\begin{equation}
\ddot{\phi}_{\mathrm{max}}= -k^2\Delta\phi-\frac{dV(\phi_{\mathrm{max}})}{d\phi}= -f(\phi_0,\Delta\phi),
\end{equation}
where 
\begin{align}\label{eq:fFunction}
f(\phi_0,\Delta\phi)= k^2\Delta\phi + \frac{dV(\phi_0+\Delta\phi)}{d\phi}~.
\end{align}
In Fig. \ref{fig:fFunction} we sketch the shape of the function $f(\phi_0,\Delta\phi)$ at fixed $\phi_0$ for different inflationary models. In the absence of inhomogeneities, $\Delta\phi=0$, $f(\phi_0,\Delta\phi)<0$ which means that $\ddot{\phi}_{\mathrm{max}}>0$, i.e. the field rolls towards the reheating minimum as expected. As we increase the amplitude of the inhomogeneities  the gradients contribute as $-\nabla^2\phi= k^2\Delta\phi>0$ which is positive definite.\footnote{Note that while it is assumed that $\phi$ has a periodic profile with wavelength $k$, this positive definiteness is general even if $\phi$ takes on a more complicated profile as $\phi_{\mathrm{max}}$ is a maximum of the profile by definition.} For small $\Delta\phi$, we can approximate the potential contributions
\begin{eqnarray}
\frac{dV(\phi_0+\Delta\phi)}{d\phi}&\approx& \frac{dV(\phi_0)}{d\phi}+\frac{d^2V(\phi_0)}{d\phi^2}\Delta\phi\nonumber\\&=&-3\sqrt{\frac{\epsilon_V}{4\pi}}H_i^2\mpl+3\eta_V H_i^2\Delta\phi\,,
\end{eqnarray}
where $\epsilon_V$ and $\eta_V$ are the usual potential slow-roll parameters evaluated at $\phi_0$. In the region of the potential that supports inflation these are small so that the contribution from gradients eventually overcomes the potential contribution. This implies that $\phi_\mathrm{max}$ is pulled back into the inflationary region as we described in \cite{Clough:2016ymm}. 
 
However, as $\Delta \phi$ increases further, expanding $dV/d\phi$ is no longer a good approximation, and the potential contribution may overcome the gradient contribution, $k^2\Delta\phi$, so that $f(\phi_0,\Delta\phi)$ may take on negative values. Suppose this occurs at an inhomogeneous amplitude of $\Delta \phi_{\mathrm{crit}}$ and correspondingly $\phi_{\mathrm{crit}} = \phi_0 + \Delta \phi_{\mathrm{crit}}$, such that a zero exists at 
\begin{equation}\label{eq:cond1}
f(\phi_0, \Delta \phi_{\mathrm{crit}})= k^2\Delta \phi_{\mathrm{crit}} +\frac{dV}{d\phi}(\phi_0 + \Delta \phi_{\mathrm{crit}})=0 ~.
\end{equation}

Then as $\Delta \phi > \Delta \phi_{\mathrm{crit}}$, the inflaton will tend to roll towards the reheating minimum at least initially, and has a greater chance of failure\footnote{In \cite{East:2015ggf}, a criterion based on the \emph{spatially average} (i.e. global) force $\langle V'(\phi)\rangle > V'(\langle \phi \rangle)$ was introduced, which led to the simple condition that inflation is likely to fail when $\Delta \phi$ reaches the end of the inflationary plateau. This is consistent with our findings. Our condition, on the other hand, relies on the balancing of the local force at the point of maximal fluctuation, and as we will see, can accurately predict the point of failure (or not) even when $\Delta \phi$ reaches beyond the end of the inflationary plateau.}

Ignoring the trivial case of $f(\phi_0,\Delta\phi)<0$, $\Delta \phi\approx 0$, where inflation is well-known to succeed, we begin with $f(\phi_0,\Delta\phi)>0$ -- we assume there exists some region where inflation will succeed as long as $\Delta \phi$ is small enough.  As we increase $\Delta \phi$, the slope of $f(\phi_0,\Delta\phi)$ is initially positive because $\eta_V\ll1$ and $k\gtrsim H_i$. We briefly described this behaviour in \cite{Clough:2016ymm}. We can further relate this to the overall shape of the potential as follows.

If the potential is convex, $d^2V/d\phi^2>0$, then $f(\phi_0,\Delta\phi)$ must increase as $\Delta \phi$ is increased and remains positive. This means that convex models are automatically robust to inhomogeneities in the inflaton sector. The inhomogeneous field is initially always pulled back, away from the reheating minimum. The time scale is roughly given by the inverse of its wavenumber, $t_{\mathrm{pb}}\sim k^{-1}$. For $N=1$ (i.e. near horizon size mode), this is then $t_{\mathrm{pb}}\sim H^{-1}$, which means that the field is pulled back within a Hubble time. We will confirm this intuition numerically in Sec. \ref{sect:results}.

On other hand, if the potential is concave, $d^2V/d\phi^2<0$, then $f(\phi_0,\Delta\phi)$ may turn over as $\Delta \phi$ increases. If the potential is sufficiently concave before the reheating minimum (which is convex by construction), then a zero crossing at $\phi_{\mathrm{crit}}$ can exists such that  $f(\phi_0,\Delta\phi_{\mathrm{crit}})=0$. In this case $f(\phi_0,\Delta\phi)<0$ at $\phi_{\mathrm{max}}>\phi_{\mathrm{crit}}$, and the inflaton will fall into the reheating minimum and end inflation. Once a sufficiently large spatial region falls into the reheating minimum, the remaining space will be dragged down by the pressure difference between the inflating plateau and the minimum, resulting in the end of inflation within a few $e$-folds. 

This discussion implies that convex potentials are generically more robust to inhomogeneities than concave potentials. We also see that decreasing the wavelength of the inhomogeneities (and hence increasing $k^2$) makes models more robust -- the most dangerous modes are the long wavelength near horizon modes, consistent with the numerical results of \cite{Clough:2016ymm}.

As we will discuss now, robustness for \emph{concave} potentials is closely related to the characteristic scale of the potential. To see this, note that in order for $f$ to vanish at $\Delta\phi_{\rm crit}$, and to turn negative for $\Delta\phi>\Delta\phi_{\rm crit}$, it must possess a maximum, i.e.  
\begin{equation}\label{eq:fprime}
\frac{\partial f}{\partial \Delta\phi} = k^2 + V''(\phi_0+\Delta\phi)\,,
\end{equation}
must have a zero for some value of $\Delta\phi$. We now consider this requirement for different models, beginning with the monomial potentials defined in~\eqn{eqn:monomial_pot}. In this case we write
\begin{eqnarray}
\frac{\partial f}{\partial \Delta\phi}&=&k^2+2p(2p-1)\frac{V(\phi_0+\Delta\phi)}{\mpl^2}\frac{\mpl^2}{(\phi_0+\Delta\phi)^2}\\
&=&k^2\left[1+\frac{6p(2p-1)}{(kL)^2}\left(\frac{\phi_{\rm max}}{\phi_0}\right)^{2p}\frac{\mpl^2}{8\pi\phi_{\rm max}^2}\right] \nonumber\,,
\end{eqnarray}
where we have used \eqn{eq:length} in going from the first to the second line. The first term proportional to $k^2$ is the gradient term, while the second is the potential term. From the denominator of the second term we again see that increasing the wavenumber suppresses the importance of the potential term relative to the gradient term, and that inflation is more robust to inhomogeneities with higher wavenumbers.  
If the potential is concave, $p<1/2$, the second term in the parentheses is negative as expected. However, since $|\phi_{\rm max}|<|\phi_0|$ and $p>0$, it is negligible until $\phi_{\rm max}$ drops well below the reduced Planck mass. If the potential is still well approximated by a power law at this point, an instability may develop (see $\phi^{2/3}$ case in Sec. \ref{sect:results}). However, the functional form assumed here only describes the potential during the inflationary period, and as the magnitude of $\phi$ decreases the potential in a single-field model must eventually turn over
and develop a minimum. This implies that for sufficiently small $|\phi|$ the potential again becomes convex. If this transition occurs before the second term becomes large, we expect the model to be robust irrespective of whether the inflationary part of the potential is convex or concave. 

We can readily extend this discussion to the other classes of models introduced in~\ref{sec:models}. For the hilltop models~(\ref{eq:ht}), we can write (again using \eqn{eq:length})
\begin{eqnarray}
\frac{\partial f}{\partial \Delta\phi}&=&k^2-n(n-1)\frac{\Lambda^4}{\mpl^2}\frac{\mpl^2}{\mu_n^2}\left(\frac{\phi_{\rm max}}{\mu_n}\right)^{n-2}\nonumber\\
&=&k^2\left[1-\frac{3n(n-1)}{(kL)^2}\frac{\mpl^2}{8\pi\mu_n^2}\left(\frac{\phi_{\rm max}}{\mu_n}\right)^{n-2}\right] \,.
\end{eqnarray}
Again the potential is only well approximated by equation~(\ref{eq:ht}) provided $\phi_{\rm max}<\mu_n$. So a maximum can only occur if $\mu_n$ is well below the reduced Planck mass, implying that models in which the characteristic scale over which the potential departs from $\Lambda^4$ is Planckian are robust even against large field excursions.

For the plateau models~(\ref{eq:bi}), defining the potential so the field rolls towards larger values of $\phi$, we can similarly write 
\begin{equation}
\frac{\partial f}{\partial \Delta\phi}=k^2\left[1-\frac{3n(n+1)}{(kL)^2}\frac{\mpl^2}{8\pi\mu_n^2}\left(\frac{\mu_n}{|\phi_{\rm max}|}\right)^{n+2}\right] \,.
\end{equation}
In this case the potential is only well approximated by~(\ref{eq:bi}) if $|\phi_{\rm max}|>\mu_n$, and as for the hilltop models, plateau models with a Planckian characteristic scale are robust against large field excursions, and the function $f$ can only change sign for models with sub-Planckian $\mu_n$. 

For models in which the potential approaches the plateau exponentially as in (\ref{eq:explateau}), but defined so that the field rolls towards the minimum in the positive $\phi$ direction 
\begin{eqnarray}
\frac{\partial f}{\partial \Delta\phi} &=& k^2 - \frac{\Lambda^4}{\mu^2}e^{\phi_{\rm max}/\mu}\nonumber\\
 &=& k^2\left[1 - \frac{3}{(kL)^2}\frac{\mpl^2}{8\pi\mu^2}e^{\phi_{\rm max}/\mu}\right]\,,\label{eqn:fstaro}
\end{eqnarray}
we see that for $\mu$ of order the reduced Planck mass, $\partial f/\partial\Delta\phi>0$ in the regime in which the potential is well approximated by (\ref{eq:explateau}), so that there is no maximum in $f$ and no zero crossing can exist for any value of $\Delta \phi$ and $\phi_0$. Starobinsky inflation is an example of this case and will be studied in the following section.

The condition defined by \eqref{eq:fFunction} is valid for the initial hyperslice. We will now show that it still broadly remains valid at a later time of the evolution.  At later times the gradients dilute due to the expansion as $\nabla^2\phi\sim k^2 \Delta\phi^2/a(t)^2$, where $a(t)\sim e^{Ht}$ is the scale factor, such that \eqref{eq:cond1} becomes
\begin{align}\label{eq:fFunction_t}
f(\phi_{\mathrm{crit}},t)\approx k^2\Delta\phi_\mathrm{crit}~ e^{-2Ht}+ \frac{dV(\phi_{\mathrm{crit}})}{d\phi},
\end{align} 
and $\Delta\phi_\mathrm{crit}$ will take different values over time.

 Hence, the robustness of a model is not only determined by the existence of a critical value $\phi_\mathrm{crit}(t=0)=\phi_0+\Delta\phi_\mathrm{crit}(t=0)$ for which the field will initially roll towards the minimum, but the field should also restore as close as possible to $\phi_0$ before crossing $\phi_\mathrm{crit}$ at later times to inflate by enough $e$-folds.  If the rate of change of $\phi_\mathrm{crit}(t)$ is small, the pull-back will have time to homogenize the field before crossing $\phi_\mathrm{crit}$, and the spacetime will inflate as in the homogeneous case (see  $D$-brane inflation in Fig.~\ref{fig:dbrane_SF} and an $\alpha$-attractor model in Fig.~\ref{fig:alpha_att}). However, if $\Delta\phi_\mathrm{crit}(t)$ decreases with $\mathcal{N}$ faster than the pull-back reduces the amplitude, the field will fall to the minimum, ending inflation (see Fig.~\ref{fig:cubic_SF}).

\section{Numerical Validation} \label{sect:results}

In this section we demonstrate the validity of our criterion by solving the equations of general relativity numerically using the numerical relativity package GRChombo \cite{Clough:2015sqa}. The metric initial conditions for the simulations are described in Sec. \ref{sect:param}, with near-horizon scale inhomogeneity as defined in \eqn{eq:fieldinitial} and \eqn{eqn:kinetic_initial}. The  evolution equations and a summary of the parameters used for the simulations (Tab. \ref{fig:table}) are shown in the Appendix. A summary video of the field evolution for different models can be found in \underline{\color{blue} \href{https://youtu.be/yk9sGuG8hdI}{this link}}.

\subsection{Convex potentials}

As discussed above, for convex potentials $\partial f/\partial\Delta\phi> 0$, so that the gradients will always pull the field back. As a concrete representative of this class, we consider
\begin{equation}
V(\phi)=\lambda \mpl^{8/3}(-\phi)^{4/3}. \label{eqn:powerlawpot}
\end{equation}

A homogeneous initial value of the field of $\phi_0=-3.26\mpl$ with $\lambda=2.57\times 10^{-14}$ would result in $100$ $e$-folds. Note that the potential as written in \eqn{eqn:powerlawpot} is only a good approximation during inflation, and the full potential is expected to be analytic at the origin. However, the details of the transition do not affect the conclusions.

\begin{figure}[tb]
\begin{center}
{\includegraphics[width=\columnwidth]{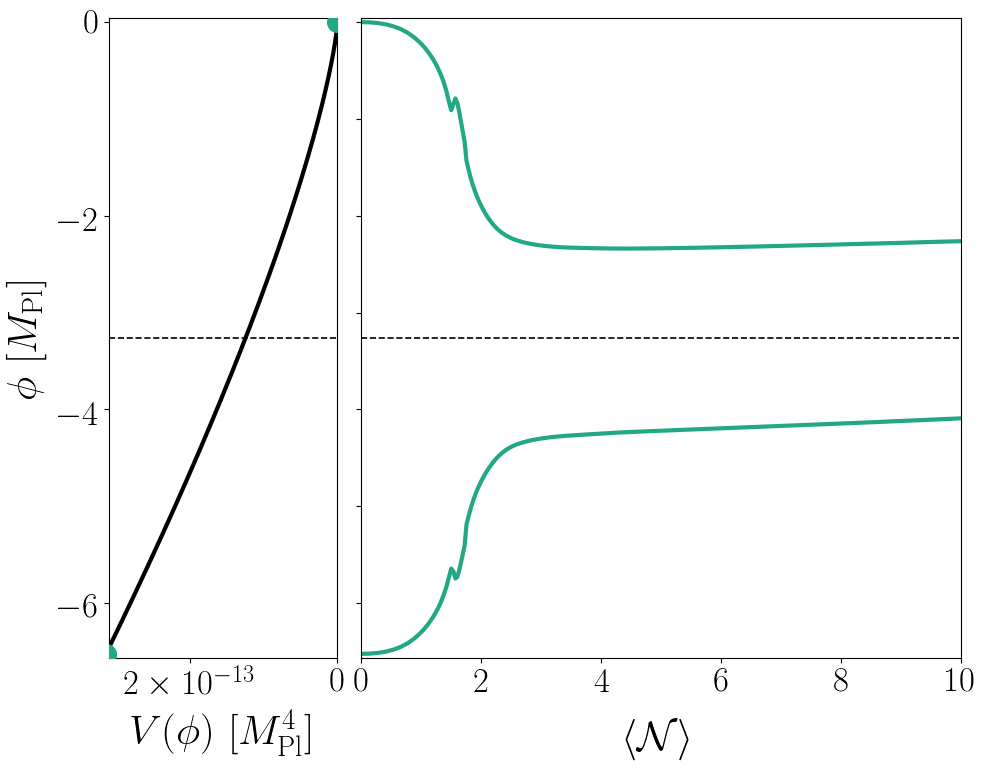}}
\caption{\textbf{Convex monomial:} The left panel shows the convex model potential $\phi^{4/3}$ (\ref{eqn:powerlawpot}) rotated by $90^\circ$. The right panel shows the evolution of the maximum $\phi_\mathrm{max}=\phi_0+\Delta\phi$ and minimum $\phi_\mathrm{min}=\phi_0-\Delta\phi$ field points as a function of $e$-folds $\mathcal{N}$ (time runs from left to right). If these extremal points restabilize to values closer to $\phi_0$, inflation can proceed. Failure occurs when one of the points diverges to the reheating minimum, but this model shows robustness to this failure mode, as expected from our analytic prescription.
}  
 \label{fig:phi43}
\end{center}
\end{figure}

Fig. \ref{fig:phi43} shows that even for field excursions that reach the reheating minimum, with $\Delta\phi=3.26\mpl$ (corresponding to $\Omega\equiv \rho_\mathrm{grad}/\rho_V \approx 50$), the maximum of the field $\phi_\mathrm{max}$ is pulled back into the inflating region, as predicted by our criterion. In the latter case, the scalar field has support well in the regime where a homogeneous field would have failed. Similar results were found for the quadratic model  $V=(1/2)m^2\phi^2$ in \cite{Clough:2016ymm}, which is also robust. 

Perhaps the most interesting point here is that convex models are robust even if the inflaton explores regions of the potential that do not support inflation. This was first observed in \cite{Clough:2016ymm} for $m^2\phi^2$ potentials, and here we see that this remains true more generally. 

It is also worth noting that different regions begin inflation for rather different values of the scalar field, so that the resulting spacetime will be highly inhomogeneous on very large scales and inflation only leads to homogeneity and isotropy within the different regions.

\subsection{Concave potentials}
As we saw in our analytic treatment, a more careful discussion is required for concave potentials because the robustness depends on the characteristic scale of the potential $\mu$. We will now test our analytic predictions for different concave models.

\subsubsection{Monomial potentials}\label{sect:phi23}

An example of a concave monomial potential which is compatible with the observed value of the spectral index $n_s$ is the so-called $\phi^{2/3}$ model which arises in the context of string theory~\cite{Silverstein:2008sg}. During inflation, the potential is well approximated by
\begin{equation}
V(\phi)=\lambda \mpl^{10/3}(-\phi)^{2/3}~. \label{eqn:phi23}
\end{equation}
This is a large field model in which $\phi_0=-2.31\mpl$ and $\lambda=3.58\times 10^{-14}$ lead to $100$ $e$-folds of inflation in the absence of inhomogeneities. As mentioned previously, for typical values of the parameters, these models develop a maximum of $f$ very close to the bottom of the potential. In fact, the critical point where $f=0$ is at $\Delta\phi_\mathrm{crit}\approx -1.25\times 10^{-10}\mpl$ for $N=1$ and thus it is challenging to simulate a region where $f<0$.

\begin{figure}[tb]
\begin{center}
{\includegraphics[width=\columnwidth]{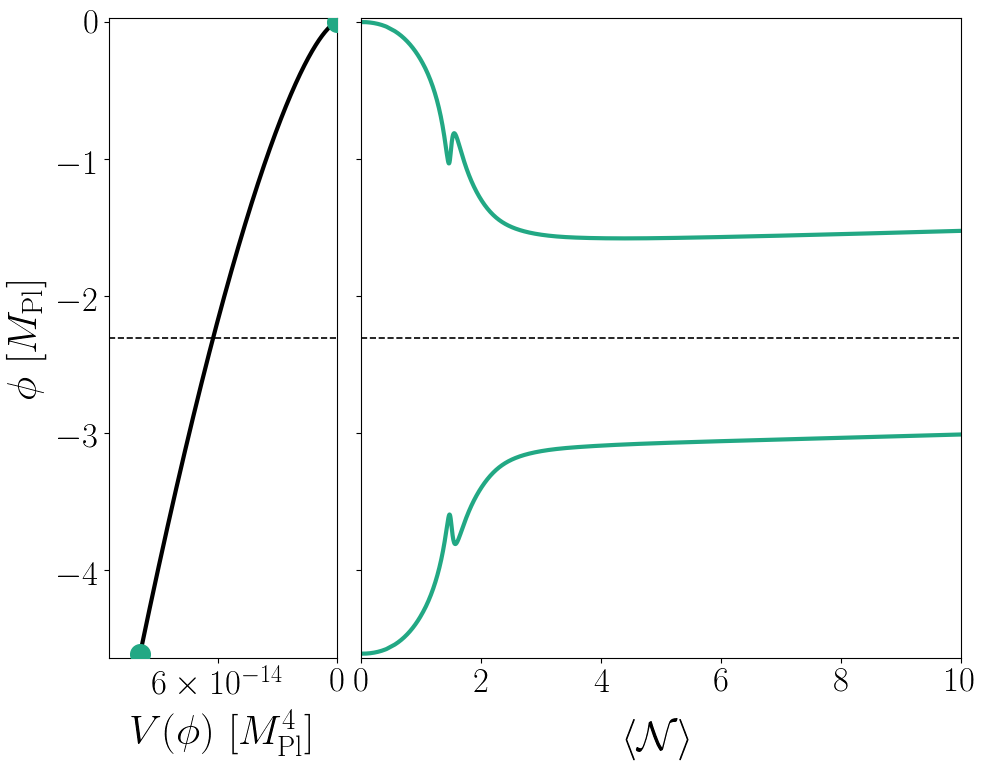}}
\caption{\textbf{Concave monomial:} As Fig. \ref{fig:phi43} but for  the concave monomial model $\phi^{2/3}$ (\ref{eqn:phi23}). The dashed black line corresponds to the mean value $\phi_0$, which is set to be the value that would result in $100$ $e$-folds for the homogeneous case. See Tab. \ref{fig:table} for more details about the parameters used. The field at the reheating minimum is pulled back and enters slow-roll inflation from the restored field value. The features at $\mathcal{N}\approx 2$ correspond to black holes that form and inflate out of the simulation grid.}
\label{fig:phi23}
\end{center}
\end{figure}

We show the results for this model in Fig.~\ref{fig:phi23}. As expected, for concave monomial models the results are similar to those with convex potentials.
 
\subsubsection{Hilltop models}

For concave models, like hilltop models~(\ref{eq:ht}), $f$ may develop a zero at $\Delta \phi_{\mathrm{crit}}$. Our condition implies that there is a maximum in $f$ if and only if $n>2$ since for $n=2$ $f$ is linear in $\Delta\phi$ and a second zero crossing cannot exist. In this section we will focus in the cubic hilltop model $n=3$,
\begin{equation}\label{eq:cubicpot}
V(\phi)=\left\{\begin{array}{cc}
 \Lambda^4 ~,&~ \phi<0 \\
 \Lambda^4\left[1-\left(\frac{\phi}{\mu_3}\right)^3\right] ~,&~ 0<\phi<\phi_{\mathrm{cc}}  \\
\frac{1}{2}m^2(\phi-\phi_\mathrm{reh})^2 ~,&~ \phi \geq \phi_{\mathrm{cc}} 
\end{array}\right.
\end{equation}
To simulate these models, we extend the inflationary potential with a quadratic minimum beyond some value $\phi_\mathrm{cc}$, and a flat plateau $V(\phi)=\Lambda^4$ for $\phi<0$. The potential we use for reheating is $V_\mathrm{reh}(\phi)=(1/2)m^2(\phi-\phi_\mathrm{reh})^2$, where $m$ and $\phi_\mathrm{reh}$ are chosen such that $V_\mathrm{inf}(\phi_\mathrm{cc})=V_\mathrm{reh}(\phi_\mathrm{cc})$ and $dV_\mathrm{inf}(\phi_\mathrm{cc})/d\phi=dV_\mathrm{reh}(\phi_\mathrm{cc})/d\phi$. The reheating potential is clearly convex, but we choose $\phi_{\mathrm{cc}}$ sufficiently deep into the non-inflating regime (i.e. the slow-roll parameter $\epsilon \geq 1$). This is conservative in the sense that the model would be more robust if the transition occurred earlier\footnote{Note that such extensions may change their observational constraints, see e.g. \cite{Kallosh:2019jnl}.}. This model of reheating is chosen for simplicity -- one could also imagine other examples like hybrid inflation where the reheating field is not the inflaton, but this would require working with an additional scalar field.

\begin{figure}[tb]
\begin{center}
{\includegraphics[width=\columnwidth]{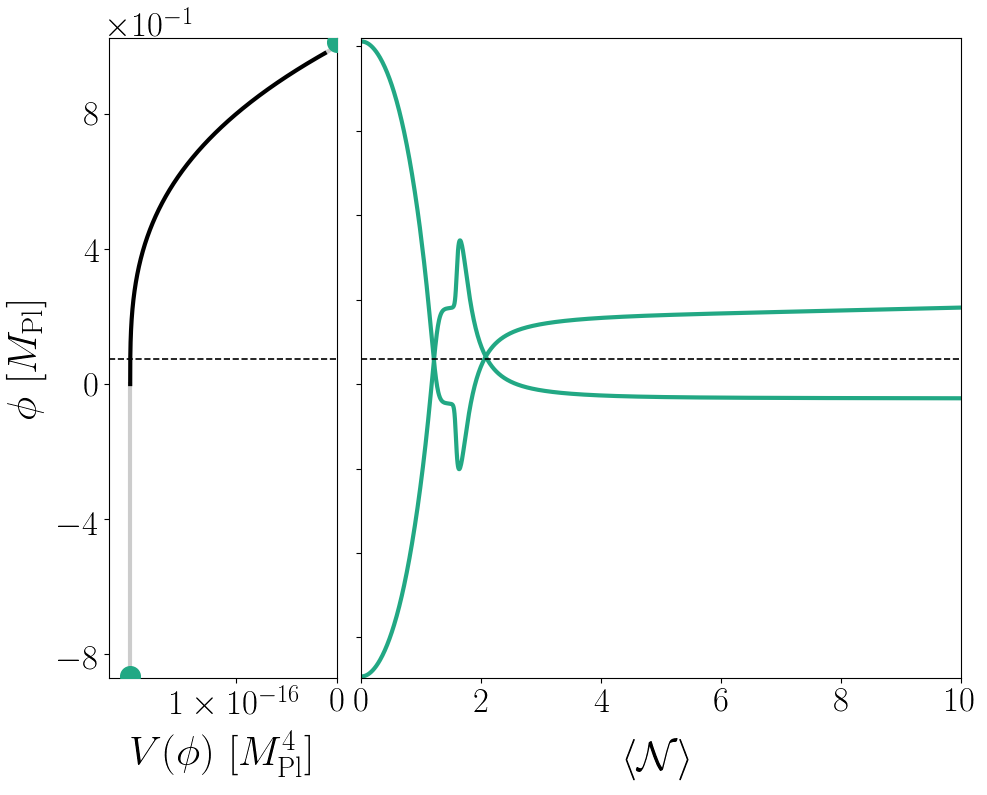}}
\caption{\textbf{Cubic hilltop} with $\mu_3=\mpl$. The left panel shows the potential (\ref{eq:cubicpot}) rotated $90^o$ where the solid black part of the curve is the inflationary potential we want to test and the grey parts correspond to extensions of the model as described in the text. The dashed black line corresponds to the mean value of the field $\phi_0$ that would result in $100$ $e$-folds in the absence of inhomogeneities. The right panel shows the evolution of the maximum and minimum of the field $\phi_\mathrm{max}$ and $\phi_\mathrm{min}$ where we have chosen $\Delta\phi$ such that the field configuration reaches the minimum of the potential. As there exists no $\Delta\phi_\mathrm{crit}$, the extrema of the field pull back towards $\phi_0$ during the evolution (blue solid line). At $\mathcal{N}\approx 2$ black holes form and shortly afterwards inflate out of the simulation when they fall below the numerical resolution of the grid.}  
\label{fig:cubic_LF}
\end{center}
\end{figure}

As discussed above, the robustness of this model depends on the characteristic scale of the potential $\mu_3$. For $\mu_3=\mpl$ we find that $\phi_0=7.43\times 10^{-2}\mpl$ and $\Lambda^4=2.05\times 10^{-16}\mpl^4$ would result in $100$ $e$-folds for the homogeneous case. As argued in the previous section, $f$ does not have a maximum, so that the field is pulled back even when it explores the minimum of the potential. This is shown in Fig. \ref{fig:cubic_LF}.

\begin{figure}[tb]
\begin{center}
{\includegraphics[width=\columnwidth]{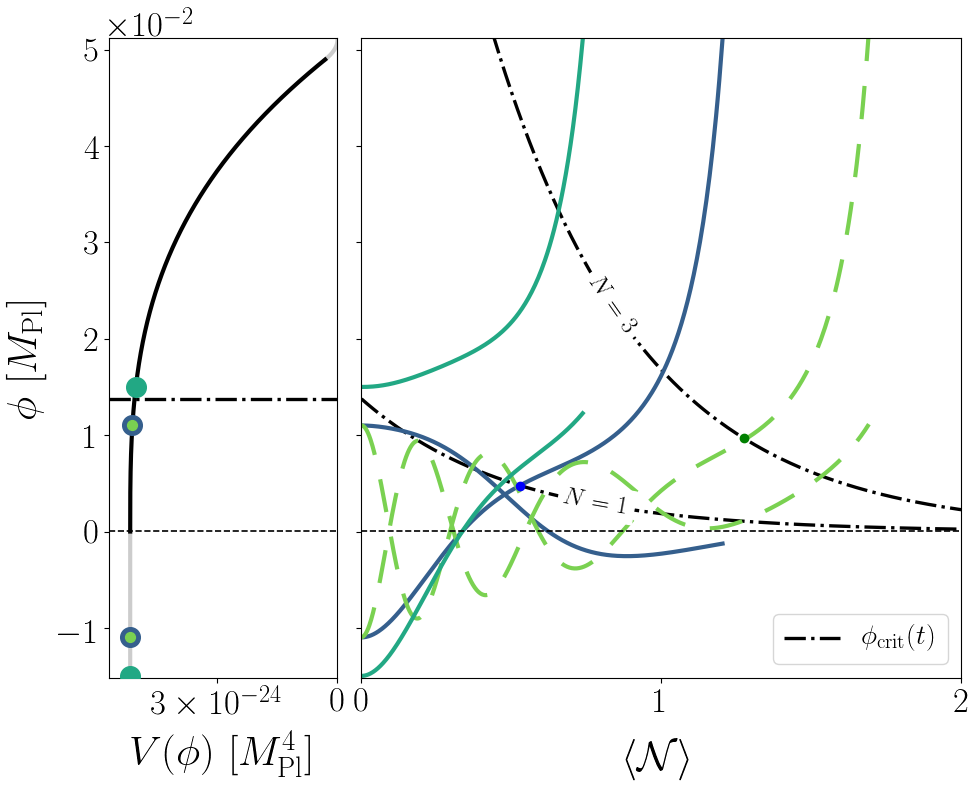}}
\caption{\textbf{Cubic hilltop} with $\mu_3=5\times 10^{-2}\mpl$. The dash-dotted black lines correspond to the critical values of the field $\phi_\mathrm{crit}(t)$ by solving \eqref{eq:fFunction_t} for different wavelengths: $k=2\pi H$ ($N=1$) and $k=6\pi H$ ($N=3$). If some value crosses to $>\phi_\mathrm{crit}(t)$, the field will fall from the inflationary plateau to the reheating minimum, dragging the rest down. For example, for $N=1$ the minimum of the field $\phi_\mathrm{min}$ (blue solid line) crosses the dash-dotted line $\phi_\mathrm{crit}$ and hence rolls down. For fixed $\Delta\phi$, the $N=3$ case (dashed green line) falls to the minimum when $\phi_\mathrm{max}$ crosses the $\phi_\mathrm{crit}$ that corresponds to $N=3$, but stays longer in the inflationary plateau than the $N=1$ case. As expected, this shows that inflation is more robust to inhomogeneities with shorter wavelengths. The parameters used in this plot are shown in Tab. \ref{fig:table}.}  
\label{fig:cubic_SF}
\end{center}
\end{figure}

On the other hand, for $\mu_3=5\times 10^{-2}\mpl$ together with $\phi_0=1.05\times 10^{-5}\mpl$ and $\Lambda^4=5.15\times 10^{-24}\mpl^4$, $f$ develops a maximum at $\Delta\phi_*=6.87\times 10^{-3}\mpl$ and a zero crossing at $\Delta\phi_\mathrm{crit}=1.38\times 10^{-2}\mpl$. So for an initial amplitude of $\Delta\phi=1.10\times 10^{-2}\mpl~<\Delta\phi_\mathrm{crit}$ the field will initially be pulled back, whereas for $\Delta\phi=1.50\times 10^{-2}\mpl~>\Delta\phi_\mathrm{crit}$, the the inflaton will fall from the inflating plateau into the reheating region, Fig. \ref{fig:cubic_SF}. 

However, even if the field is initially pulled back, inflation may fail at later times because $\phi_\mathrm{max}>\phi_\mathrm{crit}(t)$ (see dash-dotted black line). For these two cases in which $\Omega \approx 10^{-3}$, inflation fails to provide more than $1.5$ $e$-folds. In addition, we show that inflation is more robust to inhomogeneities with higher wavenumbers. Starting with the same amplitude, we see that $N=3$ leads to more $e$-folds of inflation than the $N=1$ case, although this is not sufficient to save inflation in this case. Similar results were found for the hilltop quartic model ($n=4$) in \cite{Clough:2016ymm}, in agreement with the theoretical prediction. We conclude that small field hilltop models with $n>2$ are sensitive to initial inhomogeneities in the scalar field. 

\subsubsection{Plateau models}

\begin{figure}[tb]
\begin{center}
{\vspace{0.2cm}
\includegraphics[width=\columnwidth]{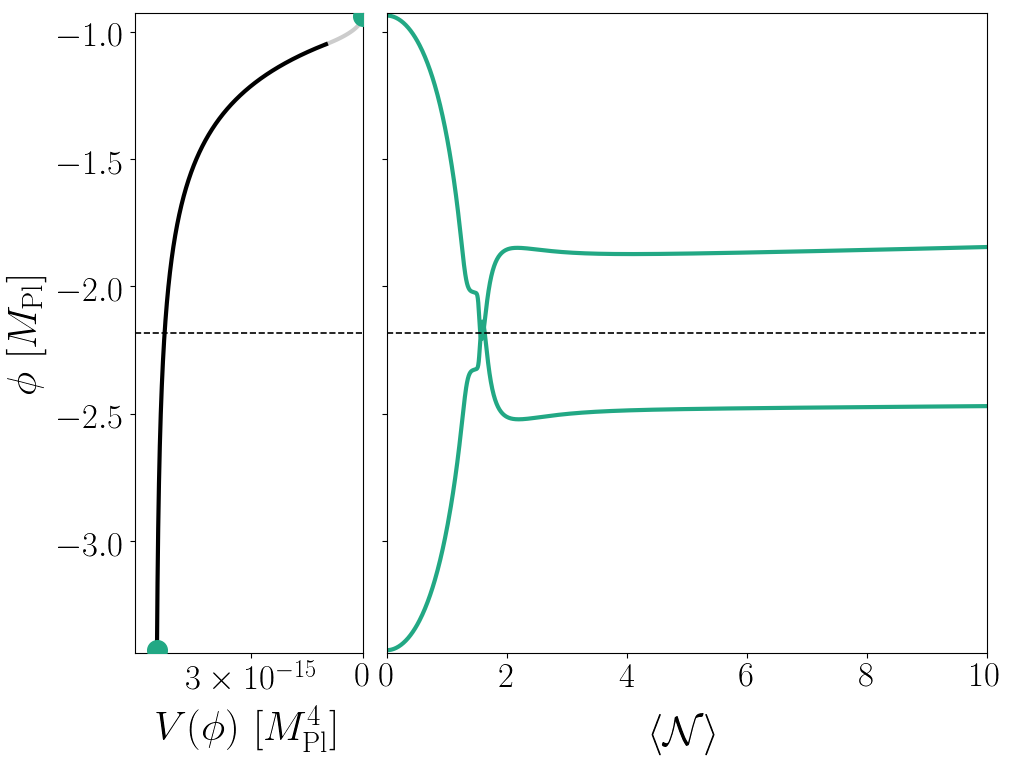}}
\caption{\textbf{Large field $D$-brane} model with $\mu_4=\mpl$. The black curve is the inflationary model and the grey parts of the curve show the reheating extension which is attached. No $\phi_\mathrm{crit}$ exists so even field configurations that reach the minimum are pulled back to the inflationary plateau. As for other large field models, the gradient energy density collapses and forms black holes at $\mathcal{N}\approx 2$. The dashed black line is the value of $\phi_0$ that would inflate for $100$ $e$-folds in the absence of inhomogeneities, see Tab. \ref{fig:table} for more details.}
\label{fig:dbrane_LF}
\end{center}
\end{figure}

We also consider the third class of concave potentials~(\ref{eq:bi}), which arises in string theory as $D$-brane inflation~\cite{Kachru:2003sx,GarciaBellido:2001ky,Dvali:2001fw}. In the best-studied case the inflaton describes the position of a $D3$ brane, which corresponds to $n=4$. As in the hilltop model, we smoothly extend the potential with a convex model at $\phi_{\mathrm{cc}}=-1.05\mu_4$ (such that  $\epsilon(\phi_{\mathrm{cc}})\gg 1$), giving us the following approximation to the potential
\begin{equation}\label{eq:dbranpot}
V=\left\{\begin{array}{cc}
 \Lambda^4\left[1-\left(\frac{\mu_4}{\phi}\right)^4\right] ~,&~ \phi<\phi_{\mathrm{cc}}  \\
\frac{1}{2}m^2(\phi-\phi_\mathrm{min})^2 ~,&~ \phi \geq \phi_{\mathrm{cc}} 
\end{array}\right.
\end{equation}
As before, we first consider $\mu_4=\mpl$, $\phi_0=-2.18\mpl$ and $\Lambda^4=5.58\times 10^{-15}\mpl^4$ so that observational constraints on the scalar power index are satisfied. We choose $\Delta\phi=1.25\mpl$ for which $\phi_\mathrm{max}=\phi_\mathrm{reh}$ and observe that the field is pulled back to safety concluding that the model is robust. This is shown in Fig.~\ref{fig:dbrane_LF}.

\begin{figure}[tb]
\begin{center}
{\includegraphics[width=\columnwidth]{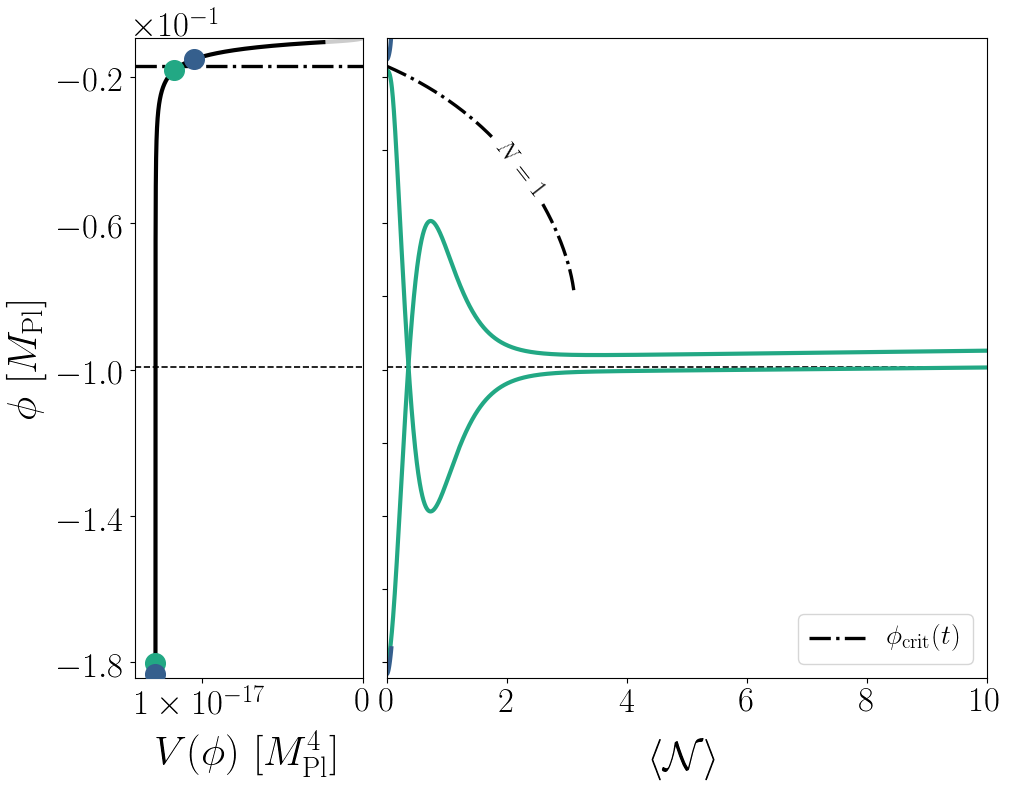}}
\caption{\textbf{Small field $D$-brane:} As Fig. \ref{fig:dbrane_LF} but with $\mu_4=10^{-2}\mpl$. For the wavenumber $N=1$, \eqref{eq:fFunction_t} predicts the field critical values $\phi_\mathrm{crit}(t)$ (dash-dotted black line). If we choose $\Delta\phi$ such that $\phi_\mathrm{max}>\phi_\mathrm{crit}$ the field will immediately fall down to the reheating minimum (blue line). The value of $\phi_\mathrm{crit}$ decrease more slowly than in the cubic hilltop case (see Fig. \ref{fig:cubic_SF}), so the field has time to pull back to values close to $\phi_0$ and slow-roll down from the plateau (green line).}
\label{fig:dbrane_SF}
\end{center}
\end{figure}

On the other hand, for $\mu_4=10^{-2}\mpl$ with $\phi_0=-9.92\times 10^{-2}\mpl$ and $\Lambda^4=1.29\times 10^{-24}\mpl^4$, we predict a value of $\Delta\phi_\mathrm{crit}=8.21\times 10^{-2}\mpl$ for which the field falls from the inflationary plateau. We test this numerically in Fig. \ref{fig:dbrane_SF} by exploring the evolution of field configurations with $\Delta\phi=8.40\times 10^{-2}\mpl$ and $\Delta\phi=8.10\times 10^{-2}\mpl$. As shown in the figure the former is immediately dragged down ending inflation while the latter is pulled back from the brink and inflates, as our criteria predicted. We conclude that small field $D$-brane inflation is not generically robust because there exist values of $\Delta\phi$ that can rapidly end it. 

The $D$-brane model is discussed further in Sec. (\ref{sect:constraints}), in the context of constraints on the initial value of $\phi_0$.

\subsubsection{Exponential plateau models} \label{sect:mixed}

For completeness, we also consider models in which the potential approaches the plateau exponentially rather than like a power law. The best-known model in this class is the Starobinsky model  \cite{STAROBINSKY1982175} 
\begin{equation}
V(\phi)=\Lambda^4\left(1-e^{\phi/\mu}\right)^2~,
\end{equation}
where $\mu \equiv \sqrt{3/16\pi}\mpl$ \cite{Starobinsky:1980te}. The transition from the concave to the convex domain of the potential occurs at $d^2V/d\phi^2(\phi_{\mathrm{cc}})=0$, or  $\phi_\mathrm{cc}=-\mu\ln{2}$ so there is no need to extend it. In this model $f$ does not possess a maximum, and the field is pulled back into the inflationary plateau even if it initially explores the minimum.

We show this in Fig. \ref{fig:starobinsky} for the parameters shown in Tab. \ref{fig:table} in the appendix. As expected, even field configurations that reach the bottom of the potential are restabilised to values closer to $\phi_0$, and we follow their evolution numerically for $\mathcal{N}>10$ $e$-folds, at which point the inhomogeneities have redshifted away and the local regions undergo slow-roll inflation. 

In the Starobinsky model the scale $\mu$ and the Planck scale share a common origin, but this is not the case for all exponential plateau models. The so-called $\alpha$-attractors \cite{Kallosh:2013hoa,Kallosh:2013yoa} are a class of models in which $\mu$ can vary over a wide range of scales. For sufficiently small values $f$ will cross zero for a large enough value of $\Delta \phi$ as we will show below.

\begin{figure}[tb]
\begin{center}
{\includegraphics[width=\columnwidth]{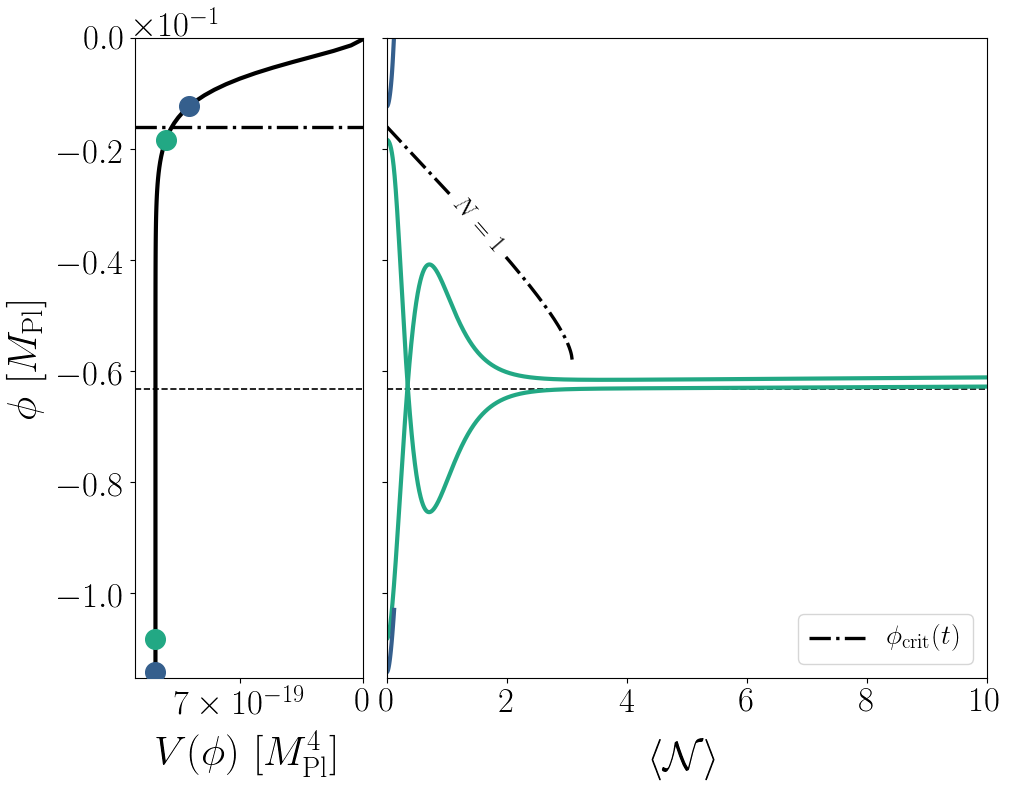}}
\caption{Example of an \textbf{$\alpha$-attractor model} with $\mu=5\times 10^{-3}\mpl$. In this case there exists a $\Delta\phi_\mathrm{crit}$ such that the field will initially roll towards the non-inflationary regime and reach the minimum (blue line). As for the small field $D$-brane model (Fig. \ref{fig:dbrane_SF}), the values taken by $\phi_\mathrm{crit}(t)$ over time change more slowly than the pull back. So as long as initially $\Delta\phi<\phi_\mathrm{crit}(t=0)$ (green line), the field will restabilise to values close to $\phi_0$ before any region crosses $\phi_\mathrm{crit}(t)$ and therefore the spacetime approaches inflation during the $10$ $e$-folds for which we follow the evolution, unlike the small field cubic hilltop model (see Fig. \ref{fig:cubic_SF}).}
\label{fig:alpha_att}
\end{center}
\end{figure}

Using \eqref{eq:fFunction} we see that for $\mu<2.7\times 10^{-2}\mpl$, $f$ will gain a maximum, and crosses zero before $\phi_\mathrm{cc}$ if $\mu< 10^{-2}\mpl$. We show this for $\mu=5\times 10^{-3}\mpl$ in Fig. \ref{fig:alpha_att}.
As before, we set $\phi_0$ to the value of the field that would result in $100$ $e$-folds in the absence of inhomogeneities. For this value $f$ crosses zero at $\Delta\phi_\mathrm{crit}=4.74\times 10^{-2}\mpl$. We confirm that for $\Delta\phi=5.1\times 10^{-2}\mpl>\Delta\phi_\mathrm{crit}$ the field is dragged down, and that it is pulled back for $\Delta\phi=4.5\times 10^{-2}\mpl<\Delta\phi_\mathrm{crit}$. Similar to the small-field $D$-brane inflation model, the rate of change of $\phi_\mathrm{crit}(t)$ over time is smaller than the pull-back, so that there is no crossing at later times and the field can restore to values close to $\phi_0$. We again follow the evolution for $\mathcal{N}>10$ $e$-folds.

\begin{figure}[tb]
\begin{center}
{\includegraphics[width=\columnwidth]{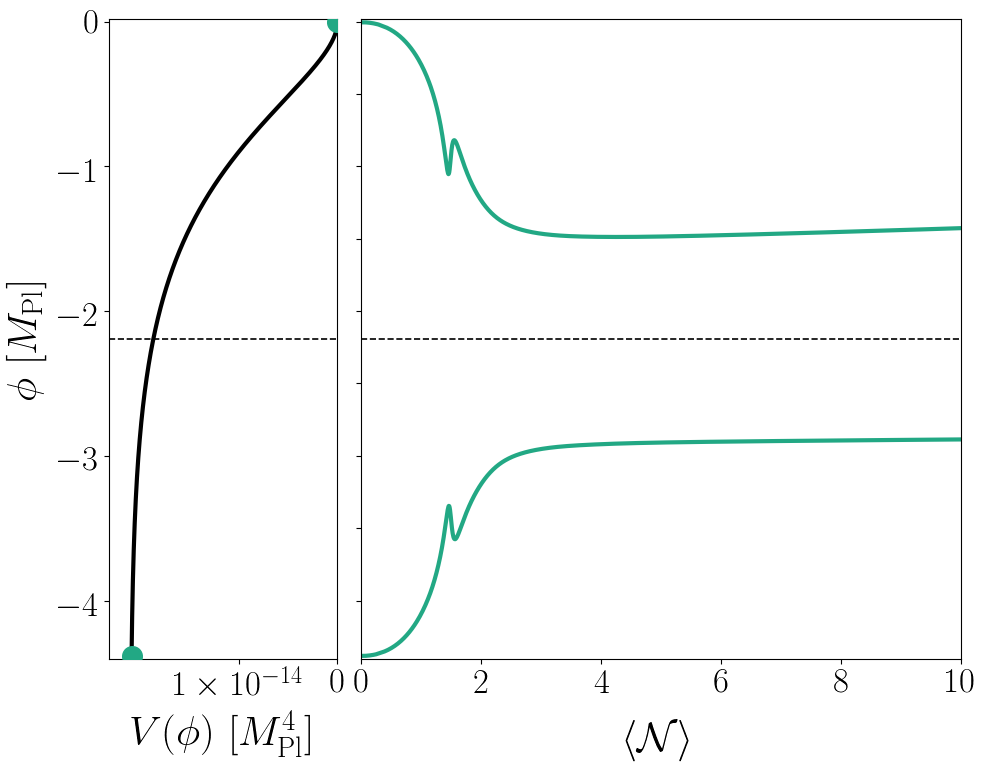}}
\caption{The \textbf{Starobinsky model} is an example of an exponential plateau model with $\mu=\sqrt{3/16\pi}\mpl$ and hence with large field excursions. As $f$ does not have a maximum, it will remain positive for any $\Delta\phi$, so that the model will support inflation even if the field configuration starts in non-inflationary regions of the potential (blue solid line). Similar to other large field cases (Fig. \ref{fig:phi23}, \ref{fig:phi43}, \ref{fig:cubic_LF} and \ref{fig:dbrane_LF}) in which there are large gradient energy densities, black holes form at $\mathcal{N}\approx 2$. See Tab. \ref{fig:table} for details of the parameters used.}
\label{fig:starobinsky}
\end{center}
\end{figure}

\section{Constraints on the Initial Value of the Scalar Field}\label{sect:constraints}

So far we have focused the discussion on models of inflation with $\phi_0$ chosen to yield $100$ $e$-folds in the absence of inhomogeneities. We will now consider both $\phi_0$ and the amplitude $\Delta\phi$ as free parameters and study the behavior of $f$ when varying them. Models that are always robust to inhomogeneities do not offer further insight into the initial value of the inflaton field $\phi_0$ required for successful inflation. However, for models that suffer the weakness of having $f<0$ for some range of $\Delta\phi$, we can obtain a constraint on $\phi_0$.

To see this, let us denote the value of $\phi$ for which $dV(\hat{\phi})/d\phi$ is the most negative by $\hat{\phi}$. This is the value of the field for which the model is most likely to fail for any value of $\Delta \phi$ as we can see from \eqn{eq:cond1}. If $f(\phi_0,\hat{\phi}-\phi_0)> 0$, the model will be robust to inhomogeneities. This condition leads to a bound 
\begin{equation}\label{eq:constraint}
\hat{\phi}-\phi_0>-\frac{1}{k^2}\frac{dV(\hat{\phi})}{d\phi}\,,
\end{equation}
that guarantees robustness for any value of $\Delta \phi$. 

Of course, not all values of $\Delta \phi$ are allowed because we must ensure that the energy density in gradients remains sub-Planckian, which imposes a bound of the form
\begin{equation}
\hat{\phi}-\phi_0\ll \frac{\mpl^2}{k}\,.
\end{equation}
 So one may ask whether one can find values of $\phi_0$ such that $f(\phi_0,\Delta\phi)> 0$ for all energetically allowed values of $\Delta\phi$, and by combining the bounds, we see that this will be the case provided
\begin{equation}
-\frac{1}{k}\frac{dV(\hat{\phi})}{d\phi}\ll\mpl^2\,.
\end{equation}

\begin{figure}[tb]
\begin{center}
{\vspace{0.06cm}
\includegraphics[width=\columnwidth]{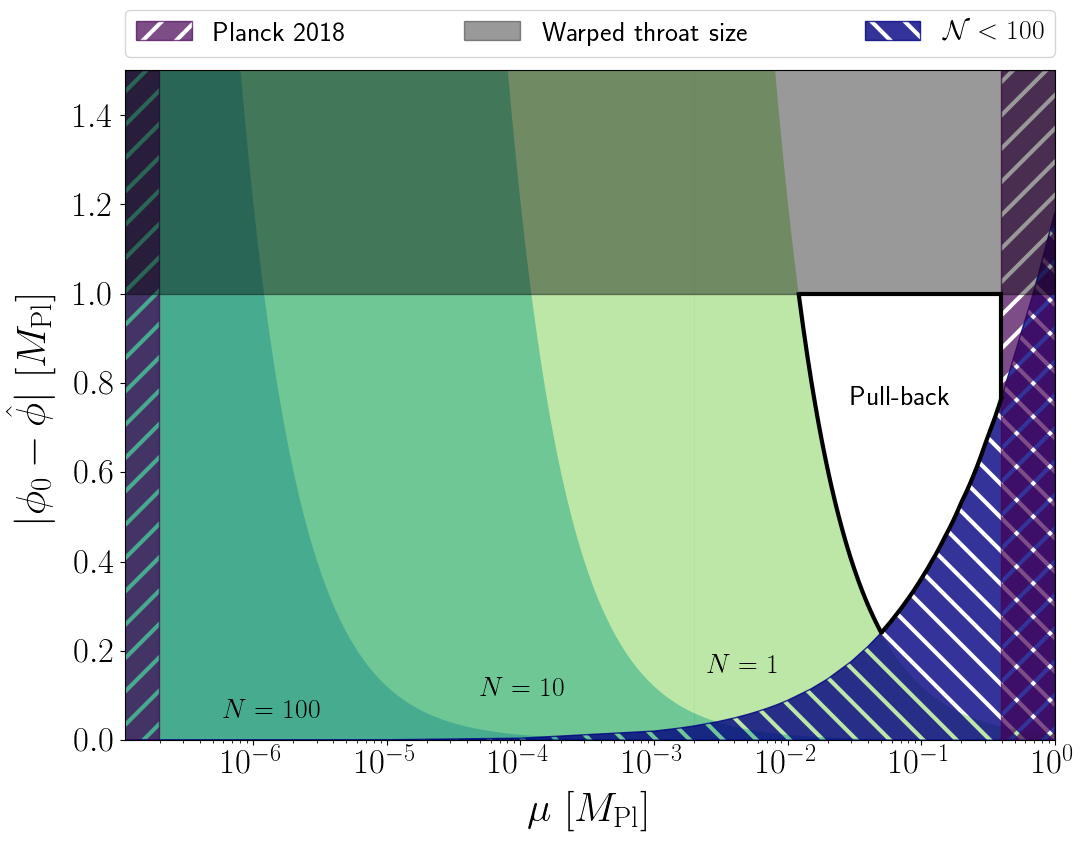}}
\caption{\textbf{Constraints on $\phi_0$ and $\mu$} for successful inflation in $D$-brane model, given by \eqref{eq:dbranpot}. We define successful inflation if a model with a mean value of $\phi_0$ has $f>0$ for all choices of $\Delta\phi$ so that only pull-back effects can be observed. Small field models are more sensitive to inhomogeneities since smaller values of $\mu$ make the model more concave and hence less robust, requiring stronger constraints for the initial mean field value $\phi_0$. Higher modes (greater N) are more robust, and therefore relax the constraints on $\phi_0$.
}
\label{fig:dbrane_constraint}
\end{center}
\end{figure}

This bound is fairly weak for any model in which the inflationary Hubble scale is well below the Planck scale, and often there will be stronger constraints on $\Delta\phi$, for example, because the potential is only of the assumed form over a much smaller field range than $\mpl^2/k$. These constraints should then be taken into account.

Let us, for example consider $D$-brane inflation. In this case, inflation must certainly end before $V(\hat{\phi})=0$, and we will take $\hat{\phi}=-\mu$ to approximate the point when $dV(\hat{\phi})/d\phi=-4\Lambda^4/\mu$ is most negative. Hence  \eqref{eq:constraint} leads to 
\begin{equation} \label{eqn:dbraneconstraint}
\phi_0<-\mu-\frac{1}{k^2}\frac{4\Lambda^4}{\mu}\,.
\end{equation}
This constraint on the initial mean value of the field is shown in Fig. \ref{fig:dbrane_constraint} for different wavenumbers of the inhomogeneities $k=2\pi N/L$ with $L=H_i^{-1}$ and $N=1$, $N=10$, and $N=100$. The green shaded regions indicate the initial conditions for which the model is not robust, in the sense that the existence of sufficiently large pertubations will cause it to fail.

In addition, we approximately sketch the constraints imposed by the $Planck$ data, which roughly exclude values of $\mu< 2\times 10^{-7}\mpl$ and $\mu> 2\times 10^{-1}\mpl$ ~\cite{Akrami:2018odb}, together with the requirement that inflation lasts for at least $\mathcal{N}=100$.

As we discussed, the energy density stored in gradients must be sub-Planckian, which imposes the additional constraint
\begin{equation}
|\phi_0|-\mu\ll\frac{\mpl^2}{k}\,.
\end{equation}
Since this bound is rather weak for typical parameters, it is not shown in Fig. \ref{fig:dbrane_constraint}. 

From the perspective of low energy effective field theory, the plateau may extend over large or possibly even infinite distances in field space. However, in the context of string theory there is another constraint we should impose on $\Delta\phi$ to ensure our discussion remains valid. The inflaton parameterizes the position of a brane along a warped throat in the internal space whose size is limited. As a consequence, the plateau only extends over a finite range, and we only expect the potential to be well-approximated by our model for sub-Planckian $\Delta\phi$~\cite{Baumann:2006cd}. So in string theory, only the area below the grey region is available, and we see that the sweet-spot for which the model is robust to inhomogeneities lies at around $\mu\sim 10^{-1}\mpl$. 

These constraints should not be taken as definitive, but rather illustrative, because other factors may influence the dynamics. For example, the presence of a sharp rise in the potential was seen to make the model less robust in~\cite{Clough:2016ymm}, and we are only studying a subset of possible inhomogeneous initial conditions. Overall, however, such differences are likely to make the constraints on the model more restrictive, rather than less.

\section{Discussion} \label{sect:discussion}

We studied the robustness of different single-field models of inflation to inhomogeneities in the scalar field. We found a simple analytic criterion that successfully predicts whether a given model for a given set of initial data will successfully inflate in our $3+1$ dimensional numerical relativity simulations. For convex potentials, we showed that inflaton eventually begins even if the inhomogeneities in initial configuration are large enough to explore the minimum of the potential. For concave models, we see that effects of inhomogeneities strongly depend on the characteristic scale of the potential. For potentials with super-Planckian characteristic scale the inflaton is pulled back towards field values for which the model supports inflation, even if the field initially explores the minimum. For potentials with sub-Planckian characteristic scales the potential gradients win over gradients in the scalar field, and inflation rapidly ends. As a consequence, concave potentials with sub-Planckian characteristic scale will require additional physical mechanisms (or tuning) to set up initial conditions suitable for inflation. 

For example, from Fig. \ref{fig:dbrane_constraint}, we see that $D$-brane inflation with $\mu<10^{-2}\mpl$ will only be robust to inhomogeneities in the field if the mean initial value is super-Planckian.  This is in tension with the bound on the field range derived in~\cite{Baumann:2006cd} and implies that inflation will not succeed if the brane is initially highly perturbed.\footnote{The brane is usually assumed to be homogeneous in most constructions of such models. See for example \cite{Brandenberger:2003py,Franche:2010yj} for studies that deviate from this assumption.}

While we have assumed that the metric sector is initially conformally flat in our analysis, since $f(\phi)\subset -dV/d\phi$, we expect the result that concave potentials with sub-Planckian characteristic scales are less robust to hold more generally. The condition \eqn{eq:fFunction} will in general contain additional curvature terms of order $\lesssim K\partial_\mu \phi$  which may change the position of the zero of $f$, but these will not dominate the $k^2\Delta \phi$ term. We leave a more detailed study of the larger space of initial conditions for future work.

\acknowledgments
We acknowledge useful conversations with  Daniel Baumann, Kieran Finn, Thomas Helfer, Cristian Joana, Sotirios Karamitsos.  
KC acknowledges funding from the European Research Council (ERC) under the European Union’s Horizon 2020 research and innovation programme (grant agreement No 693024).
RF is supported in part by the Alfred P. Sloan Foundation, the Department of Energy under grant de-sc0009919, and the Simons Foundation/SFARI 560536.
EAL is supported by an STFC AGP-AT grant (ST/P000606/1) and a FQXi Large Grant (``\emph{Minimal Observers and Maximal Observations}''). 
We would also like to thank the GRChombo team
\href{http://www.grchombo.org}{(http://www.grchombo.org/)} and the COSMOS team at DAMTP, Cambridge University for their ongoing technical support.
This work was performed using the Cambridge Service for Data Driven Discovery (CSD3), part of which is operated by the University of Cambridge Research Computing on behalf of the STFC DiRAC HPC Facility (www.dirac.ac.uk). The DiRAC component of CSD3 was funded by BEIS capital funding via STFC capital grants ST/P002307/1 and ST/R002452/1 and STFC operations grant ST/R00689X/1. DiRAC is part of the National e-Infrastructure.

\bibliography{convexconcave.bib}
\clearpage 
 
 \section{Appendix}

\subsection{Evolution Equations}
In this work, we use \textsc{GRChombo}, a multipurpose numerical relativity code \cite{Clough:2015sqa} which solves the BSSN \cite{Baumgarte:1998te,PhysRevD.52.5428,Shibata:1995we} formulation of the Einstein equation. The 4 dimensional spacetime metric is decomposed into a spatial metric on a 3 dimensional spatial hypersurface, $\gamma_{ij}$, and an extrinsic curvature $K_{ij}$, which are both evolved along a chosen local time coordinate $t$. The line element of the decomposition is
\begin{equation}
ds^2=-\alpha^2\,dt^2+\gamma_{ij}(dx^i + \beta^i\,dt)(dx^j + \beta^j\,dt)~,
\end{equation}
where $\alpha$ and $\beta^i$ are the lapse and shift, gauge parameters.  These gauge parameters are specified on the initial hypersurface and then allowed to evolve using gauge-driver equations, in accordance with the puncture gauge \cite{Campanelli:2005dd,Baker:2005vv}, for which the evolution equations are
\begin{align} \label{eqn:MovingPuncture}
&\partial_t \alpha = - \mu \alpha K + \beta^i \partial_i \alpha ~ , \\
&\partial_t \beta^i = B^i ~ , \\
&\partial_t B^i = \frac{3}{4} \partial_t \Gamma^i - \eta B^i ~ ,
\end{align}
where the constants $\eta$ and $\mu$ are of order $1/M_{\mathrm{ADM}}$ and unity respectively.  

The induced metric is decomposed as 
\begin{equation}\label{eq:metric}
\gamma_{ij}=\frac{1}{\chi}\tilde\gamma_{ij}~,~\det\tilde\gamma_{ij}=1~,~ \chi = \left(\det\gamma_{ij}\right)^{-\frac{1}{3}}  ~ .
\end{equation}
The extrinsic curvature is decomposed into its trace, $K=\gamma^{ij}\,K_{ij}$, and its traceless part $\tilde\gamma^{ij}\,\tilde A_{ij}=0$ as
\begin{equation}
K_{ij}=\frac{1}{\chi}\left(\tilde A_{ij} + \frac{1}{3}\,K\,\tilde\gamma_{ij}\right) ~ .
\end{equation}
The conformal connections are $\tilde\Gamma^i=\tilde\gamma^{jk}\,\tilde\Gamma^i_{~jk}$ where $\tilde\Gamma^i_{~jk}$ are the Christoffel symbols associated with the conformal metric $\tilde\gamma_{ij}$.
The evolution equations for the gravity sector of BSSN are then
\begin{align}
&\partial_t\chi=\frac{2}{3}\chi \alpha K-\frac{2}{3}\chi \partial_k\beta^k+\beta^k\partial_k\chi ~ , \label{eqn:dtchi2} \\
&\partial_t\tilde\gamma_{ij} =-2\,\alpha\, \tilde A_{ij}+\tilde \gamma_{ik}\,\partial_j\beta^k+\tilde \gamma_{jk}\,\partial_i\beta^k \nonumber \\
&\hspace{1.3cm} -\frac{2}{3}\,\tilde \gamma_{ij}\,\partial_k\beta^k +\beta^k\,\partial_k \tilde \gamma_{ij} ~ , \label{eqn:dttgamma2} \\
&\partial_t K = -\gamma^{ij}D_i D_j \alpha + \alpha\left(\tilde{A}_{ij} \tilde{A}^{ij} + \frac{1}{3} K^2 \right) \nonumber \\
&\hspace{1.3cm} + \beta^i\partial_iK + 4\pi\,\alpha(\rho + S) \label{eqn:dtK2} ~ , \\
&\partial_t\tilde A_{ij} = \chi\left[-D_iD_j \alpha + \alpha\left( R_{ij} - 8\pi\,\alpha \,S_{ij}\right)\right]^\textrm{TF} \nonumber \\
&\hspace{1.3cm} + \alpha (K \tilde A_{ij} - 2 \tilde A_{il}\,\tilde A^l{}_j)  \nonumber \\
&\hspace{1.3cm} + \tilde A_{ik}\, \partial_j\beta^k + \tilde A_{jk}\,\partial_i\beta^k \nonumber \\
&\hspace{1.3cm} -\frac{2}{3}\,\tilde A_{ij}\,\partial_k\beta^k+\beta^k\,\partial_k \tilde A_{ij}\,   \label{eqn:dtAij2} ~, \\ 
&\partial_t \tilde \Gamma^i=2\,\alpha\left(\tilde\Gamma^i_{jk}\,\tilde A^{jk}-\frac{2}{3}\,\tilde\gamma^{ij}\partial_j K - \frac{3}{2}\,\tilde A^{ij}\frac{\partial_j \chi}{\chi}\right) \nonumber \\
&\hspace{1.3cm} -2\,\tilde A^{ij}\,\partial_j \alpha +\beta^k\partial_k \tilde\Gamma^{i} \nonumber \\
&\hspace{1.3cm} +\tilde\gamma^{jk}\partial_j\partial_k \beta^i +\frac{1}{3}\,\tilde\gamma^{ij}\partial_j \partial_k\beta^k \nonumber \\
&\hspace{1.3cm} + \frac{2}{3}\,\tilde\Gamma^i\,\partial_k \beta^k -\tilde\Gamma^k\partial_k \beta^i - 16\pi\,\alpha\,\tilde\gamma^{ij}\,S_j ~ . \label{eqn:dtgamma2}
\end{align} 

Meanwhile, the matter part of the Lagrangian is 
\begin{equation}
\mathcal{L}_\phi=-\frac{1}{2}g^{\mu\nu}\partial_\mu\phi\partial_\nu\phi-V(\phi)~,
\end{equation}
which gives the evolution equations 
\begin{equation}
-\nabla_\mu \nabla^\mu\phi+\frac{d V(\phi)}{d \phi}=0~,
\end{equation}
and decomposing the matter equation into two first order equations, with BSSN variables it becomes
\begin{align}
&\partial_t \phi = \alpha \Pi_{M} +\beta^i\partial_i \phi \label{eqn:dtphi2} ~ , \\
&\partial_t \Pi_{M}=\beta^i\partial_i \Pi_{M} + \alpha\partial_i\partial^i \phi + \partial_i \phi\partial^i \alpha \nn
&\hspace{1.3cm} +\alpha\left(K\Pi_{M}-\gamma^{ij}\Gamma^k_{ij}\partial_k \phi-\frac{dV}{d\phi}\right)~.
\end{align} 
The stress energy tensor is
\begin{equation}
T_{ab} = 
\nabla_{a}\phi\nabla_{b}\phi
-\frac{1}{2}g_{ab}\left(\nabla_c\phi\nabla^c \phi+2V\right),
\end{equation}
and its various components are defined as
\begin{align}
&\rho = n_a\,n_b\,T^{ab}\,,\quad S_i = -\gamma_{ia}\,n_b\,T^{ab}\,, \nonumber \\
&S_{ij} = \gamma_{ia}\,\gamma_{jb}\,T^{ab}\,,\quad S = \gamma^{ij}\,S_{ij} ~.
\label{eq:Mattereqns}
\end{align}
The Hamiltonian constraint 
\begin{equation}\label{eq:HamiltonianConstraint}
\mathcal{H} = R + K^2-K_{ij}K^{ij}-16\pi \rho ~ ,
\end{equation}
and the momentum constraint 
\begin{equation}\label{eq:MomentumConstraint}
\mathcal{M}_i = D^j (\gamma_{ij} K - K_{ij}) - 8\pi S_i ~,
\end{equation}
are monitored throughout the evolution to check the quality of our simulations, see Fig. \ref{fig:L2H}. 
We use periodic boundary conditions in all directions.

\begin{table*}[tb]
\centering
\begin{tabular}{|c|c|c|c|c|c|c|c|c|c|}
\hline
Model $V(\phi)$            & $\quad\mu~[\mpl]\quad$   & $\lambda~ \vert ~\Lambda^4~[\mpl^4]$               & $\phi_0~[\mpl]$  & $\quad H_\mathrm{inf}~[\mpl]\quad$ 
& $\Delta\phi_\mathrm{e}~[\mpl]$ & $\Delta\phi_\mathrm{crit}~[\mpl]$ & $\Delta\phi_1~[\mpl]$ & $\Delta\phi_2~[\mpl]$ \\ \hline
$\lambda \mpl^{8/3}(-\phi)^{4/3}$     & $---$     & $2.57\times 10^{-14}$   & $-3.26$ & $1.02\times 10^{-6}$  & $3.07$  & $---$    &
 $3.26$ & $---$                   \\ \hline
$\lambda \mpl^{10/3}(-\phi)^{2/3}$ & $---$  & $3.58\times 10^{-14}$       & $-2.31$ 
&  $7.23\times 10^{-7}$  & $2.21$ & $---$       & $2.31$ & $---$               \\ \hline
\multirow{ 2}{*}{$\Lambda^4\left(1-\left(\frac{\phi}{\mu_3}\right)^3\right)$} & $1$ & $2.05\times 10^{-16}$  & $7.43\times 10^{-2}$ & $4.14\times 10^{-8}$ & $8.03\times 10^{-1}$ & $---$ & $9.40\times 10^{-1}$ & $---$ \\ 
& $5\times 10^{-2}$ & $5.15\times 10^{-24}$  & $1.05\times 10^{-5}$
& $6.57\times 10^{-12}$ & $1.68\times 10^{-2}$ & $1.38\times 10^{-2}$ & $1.10\times 10^{-2}$ & $1.50\times 10^{-2}$\\ \hline
\multirow{ 2}{*}{$\Lambda^4\left(1-\left(\frac{\phi}{\mu_4}\right)^{-4}\right)$} & $1$ & $5.58\times 10^{-15}$ & $-2.18$    
& $2.11\times 10^{-7}$  & $1.07$ & $---$              & $1.25$ & $---$  \\
& $1\times 10^{-2}$ & $1.29\times 10^{-17}$ & $-9.92\times 10^{-2}$    
& $1.04\times 10^{-8}$  & $7.67\times 10^{-2}$ & $8.21\times 10^{-2}$     & $8.10\times 10^{-2}$ & $8.40\times 10^{-2}$ \\ \hline
\multirow{ 2}{*}{$\Lambda^4\left(1-\exp\left[{\phi/\mu}\right]\right)^2$} & $\sqrt{3/16\pi}$ & $2.11\times 10^{-14} $ & $-2.19$    
& $3.97\times 10^{-7}$  & $1.95$ & $---$              & $2.19$ & $---$
\\ 
& $5\times 10^{-3}$ & $1.18\times 10^{-18} $ & $-6.33\times 10^{-2}$    
& $3.14\times 10^{-9}$  & $4.31\times 10^{-2}$ & $4.74\times 10^{-2}$              & $4.50\times 10^{-2}$ & $5.10\times 10^{-2}$ \\ \hline
\end{tabular}
 \caption{\textbf{Overview of runs:} (i) Convex monomial. (ii) Concave monomial. (iii) Cubic hilltop. (iv) $D$-brane. (v) $\alpha$-attractor model.  In these cases $\Lambda^4$ is chosen such that it is compatible with scalar index measurements from the Planck 2018 observations. In addition, $\phi_0$ is the initial field value that would correspond to $100$ $e$-folds in the absence of inhomogeneities $\Delta\phi=0$. $\Delta\phi_\mathrm{e}$ corresponds to the value where $\epsilon_V(\phi_\mathrm{e})=1$ with $\phi_\mathrm{e}=\phi_0+\Delta\phi_\mathrm{e}$. $\Delta\phi_\mathrm{crit}$ is the value for which $f(\phi_0,\Delta\phi_\mathrm{crit})=0$ and $\Delta\phi_1$ and $\Delta\phi_2$ are different amplitudes used in simulations.}  
\label{fig:table}
\end{table*}

\subsection{Initial Data}

We have defined the conformal metric $\tilde{\gamma}_{ij}=\chi\gamma_{ij}$ where $\chi$ is the conformal factor, a scalar density, as per \eqref{eq:metric}. We make the simplifying assumption that
\begin{equation}
A_{ij}\equiv K_{ij} - \frac{1}{3}\gamma_{ij}K=0
\end{equation}
and that the induced metric is conformally flat
\begin{equation}
\tilde{\gamma}_{ij}=\delta_{ij}~.
\end{equation}

We solve the Hamiltonian constraint \eqref{eq:HamiltonianConstraint} for $\chi$ using a multigrid solver. As explained in Sec. \ref{sect:param}, the momentum constraints \eqref{eq:MomentumConstraint} are trivially satisfied, and the constant value of $K$ is set by imposing an integrability condition on the periodic domain.

\subsection{Measurement of e-folds}

The number of $e$-folds informs by how much the universe has expanded from a reference time $t_0$. In an inflationary spacetime the scale factor grows as $a(t)\propto e^{Ht}$ and the number of $e$-folds $\mathcal{N}$ with respect to $t_0$ is then defined as
\begin{equation}
\mathcal{N} = \ln\left(\frac{a(t)}{a(t_0)}\right)
\end{equation}
In our code, the conformal factor is related to the scale factor in a FLRW spacetime as $\chi=a(t)^{-2}$ so that the local number of $e$-folds can be obtained by evaluating 
\begin{equation}
\mathcal{N}=-\frac{1}{2}\ln \chi~.
\end{equation}
In this work we track the average number of $e$-folds over the simulation box with coordinate volume $V = dx dy dz$ by averaging $\chi$, so that
\begin{equation}
\langle \mathcal{N} \rangle= -\frac{1}{2}\ln \langle\chi\rangle
\end{equation}
where
\begin{equation}
\langle \chi \rangle = \frac{1}{V} \int_V \chi ~dV~.
\end{equation}

\begin{figure}[tb]
\begin{center}
{\includegraphics[width=1.0\columnwidth]{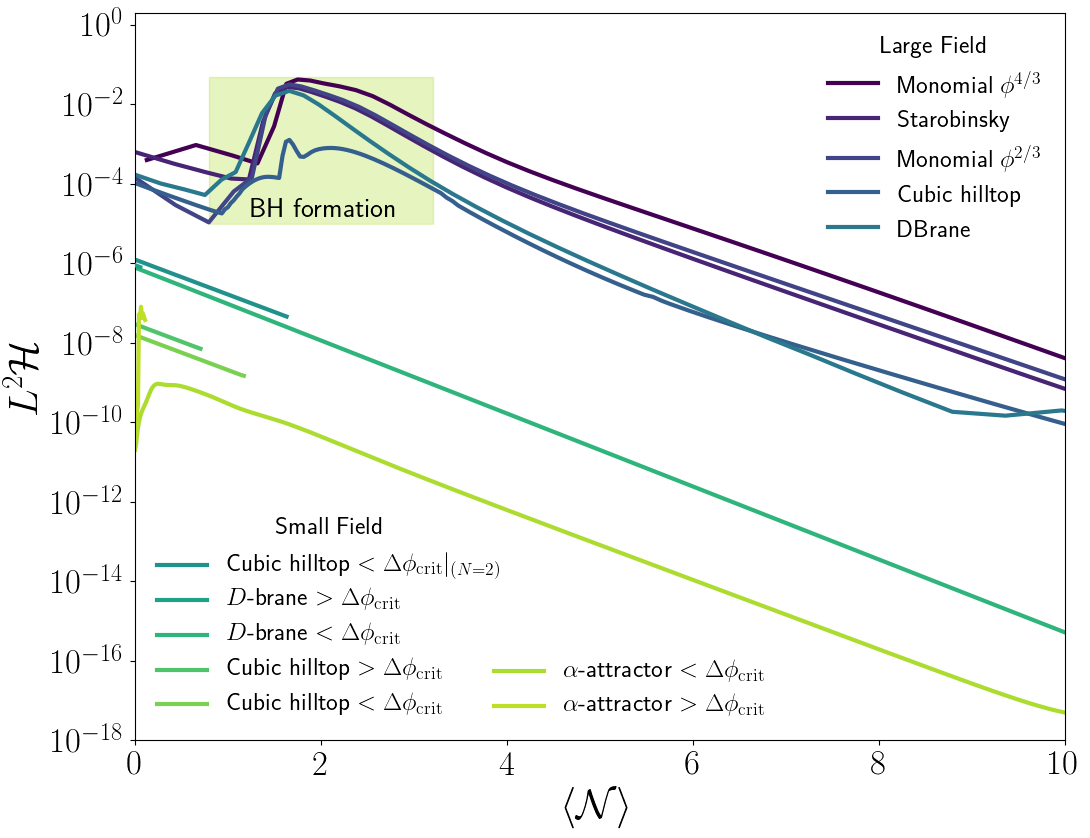}}
\vspace{-1.5em} \caption{\textbf{$L^2\mathcal{H}$ constraint} remains stable throughout the evolution for all runs. Dark colours correspond to large field models, for which black holes form at $\mathcal{N}\approx 2$ (yellow-coloured region). Lighter colours represent the Hamiltonian constraint violation for small field cases, in which the gradient energy density is not enough to form singularities. For these cases the $L^2\mathcal{H}$ analysis is stopped when the field first reaches the minimum.}
\label{fig:L2H}
\end{center}
\end{figure}

\subsection{Constraint violation}

In Fig. \ref{fig:L2H}, we show that the  volume-averaged Hamiltonian constraint violation
\begin{equation}\label{eq:L2H}
L^2\mathcal{H} =\sqrt{\frac{1}{V} \int_V |\mathcal{H}^2| dV}~,
\end{equation}
where $V$ is the box volume, is under control throughout the simulations studied in this paper.

We use the gradient conditions on $\phi$ and $\chi$ to tag cells for regridding, although in many of our simulations a single level is sufficient. It is only for the large field cases in which $\Delta\phi>\mpl$, where large gradient energies are present, that we need use AMR to resolve any collapse to black holes. As we do not excise the interior of the black holes, an increase in $L^2\mathcal{H}$ can be seen in the yellow-coloured region until the black holes are inflated out due to the expanding spacetime. In addition, cases in which the field initially rolls-down to the reheating minimum and drags-down the rest of the field, sharp gradients have to be resolved by using multiple levels of AMR which is challenging numerically due to the two different scales that need to be tracked.

\subsection{Convergence testing}

\begin{figure}[t]
\begin{center}
{\includegraphics[width=1.0\columnwidth]{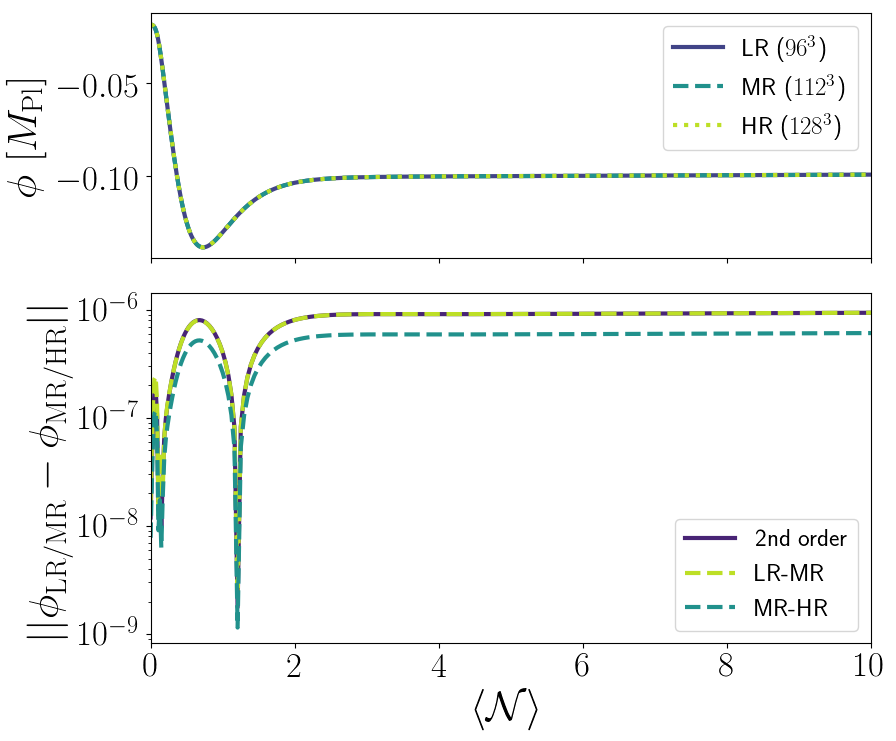}}
\vspace{-1.5em} \caption{\textbf{Convergence test} for small field $D$-brane inflation model is consistent with a 2nd order convergence. Top panel: Evolution of $\phi_\mathrm{max}$ for low (LR: $96^3$), mid (MR: $112^3$) and high (HR: $128^3$) resolutions. Bottom panel: LR-MR and MR-HR errors and the LR-MR values expected at 2nd order convergence. }
\label{fig:convergence}
\end{center}
\end{figure}

We tested the convergence of our simulations using a box of size $L=32M=H^{-1}$ with low (LR: $96^3$), mid (MR: $112^3$) and high (HR: $128^3$) resolutions. We extract the value of the field at the center of the grid, which corresponds to $\phi_\mathrm{max}$ and track its evolution. In particular we look at the case of $\Delta\phi<\Delta\phi_\mathrm{crit}$ for the small field $D$-brane model \ref{fig:dbrane_SF}. The evolution of the field with respect to the average number of $e$-folds is shown in the top panel of Fig. \ref{fig:convergence}, together with the relative errors for different resolutions in dashed lines $\vert\vert \phi_\mathrm{LR}-\phi_\mathrm{MR}\vert\vert$ and $\vert\vert \phi_\mathrm{MR}-\phi_\mathrm{HR}\vert\vert$. The evolution is consistent with 2nd order convergence (solid line).

\end{document}